\documentclass[10pt]{IEEEtran}%2-column
\usepackage{cite}
\usepackage{amsmath,amssymb,amsfonts}
\usepackage{algorithm, algorithmic}
\usepackage{graphicx}
\usepackage{textcomp}
\def\BibTeX{{\rm B\kern-.05em{\sc i\kern-.025em b}\kern-.08em
    T\kern-.1667em\lower.7ex\hbox{E}\kern-.125emX}}
\usepackage{xcolor} %when you want to color text in a document

\usepackage{subcaption}

\usepackage{authblk}
\usepackage{comment}
\usepackage{amsthm}
\usepackage{pbox}

\usepackage{array}
\usepackage{url} %Put this to cite a URL in BibTeX
\usepackage{enumitem} %a,b,c enumerating
\newcolumntype{P}[1]{>{\centering\arraybackslash}p{#1}}
\newcolumntype{M}[1]{>{\centering\arraybackslash}m{#1}}
\usepackage[utf8]{inputenc}
\usepackage[T1]{fontenc}
\usepackage{booktabs, multirow, bigstrut}%multi row
\usepackage{siunitx}
\usepackage[flushleft]{threeparttable} %Tablefootnotes
\usepackage{pifont} %Thick X-mark
\usepackage[T1]{fontenc} %Thick X-mark
\usepackage{makecell, caption} %Thick horizontal line of Table
\usepackage{rotating, graphicx} %Table: Horz and vert alignment, and % to display tables in landscape
\usepackage{makecell} %add a forced line break for a single cell

\usepackage{longtable} % Multi-page or longtables
\usepackage{academicons} %ORCID
\usepackage{tablefootnote}
\begin{document}
\title{Comprehensive Survey and Taxonomies of False Injection Attacks in Smart Grid: Attack Models, Targets, and Impacts}
\author{Haftu Tasew Reda$^{1}$, Adnan Anwar$^2$, and Abdun Mahmood$^1$ \\ 
$^1$ Department of Computer Science and IT, La Trobe University, VIC 3086, Australia. \\
$^2$ School of Information Technology, Deakin University, VIC 3216, Australia. \\}
 \maketitle
\begin{abstract}
Smart Grid has rapidly transformed the centrally controlled power system into a massively interconnected cyber-physical system that benefits from the revolutions happening in the communications (e.g. 5G) and the growing proliferation of the Internet of Things devices (such as smart metres and intelligent electronic devices). 
While the convergence of a significant number of cyber-physical elements has enabled the Smart Grid to be far more efficient and competitive in addressing the growing global energy challenges, it has also introduced a large number of vulnerabilities culminating in violations of data availability, integrity, and confidentiality. Recently, false data injection (FDI) has become one of the most critical cyberattacks, and appears to be a focal point of interest for both research and industry. To this end, this paper presents a comprehensive review in the recent advances of the FDI attacks, with particular emphasis on 1) adversarial models, 2) attack targets, and 3) impacts in the Smart Grid infrastructure. This review paper aims to provide a thorough understanding of the incumbent threats affecting the entire spectrum of the Smart Grid. Related literature are analysed and compared in terms of their theoretical and practical implications to the Smart Grid cybersecurity. In conclusion, a range of technical limitations of existing false data attack research is identified, and a number of future research directions is recommended.         
\end{abstract}
\begin{IEEEkeywords}
Smart Grid, cybersecurity, power system reliability, cyber-physical system, cyberattack, false data injection.  
\end{IEEEkeywords}

\section{Introduction} \label{sec:introduction}
\IEEEPARstart{T}{HE} major threat to critical infrastructure from nation states and hostile actors raises real challenges in identifying the operational vulnerabilities of the Smart Grid, as well as the various attack vectors that could jeopardise the reliability and performance of the power system. 

\begin{figure}[H]
	\centerline{\includegraphics[width=9cm]{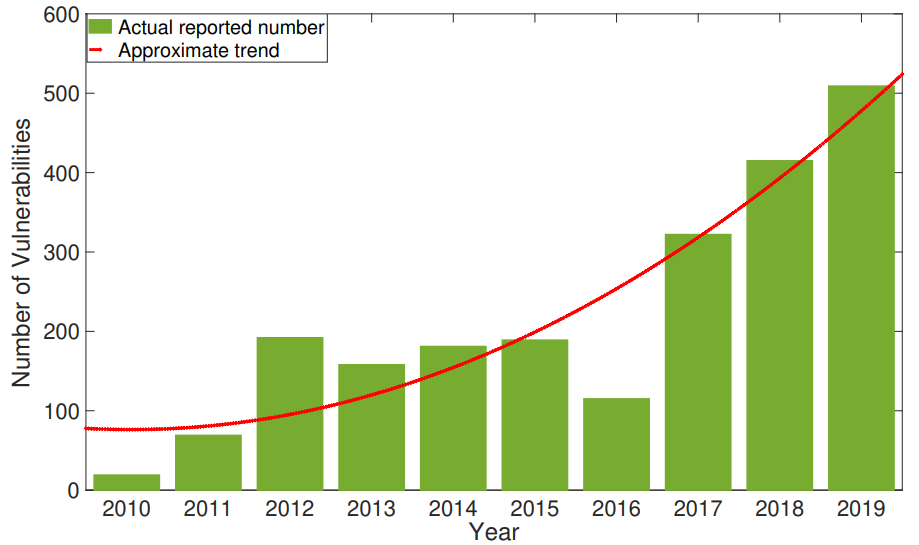}}
	\caption{Number of ICS vulnerabilities by year (reproduced from the US ICS-CERT \cite{11} and Kaspersky ICS-CERT \cite{13})}
	\label{vulEnergy}
\end{figure}

According to vulnerability reports from the US ICS-CERT \cite{11} and Kaspersky ICS-CERT \cite{13}, the energy sector has reported the greatest number of vulnerabilities among all network infrastructures. For example, Fig. \ref{vulEnergy} shows the number of vulnerabilities of various Industrial Control System (ICS) elements between 2010 and 2019 \cite{11} \cite{13}. Accordingly, 178, 110, and 283 cyberattack incidents were recorded in the energy sector out of 322, 415, and 509 ICS cyberattack incidents, respectively across the fiscal years 2017, 2018, and 2019. These cyber incidents may lead to myriads of security risks including the loss of critical data necessary for control operations, malicious modification of critical power system states. Possible consequences include incorrect customer billing information, price manipulation in the energy market, small to large scale electric power outage, and the likelihood of endangering lives by limiting power to other national critical infrastructures.

\begin{figure}[H]
	\centerline{\includegraphics[width=9cm]{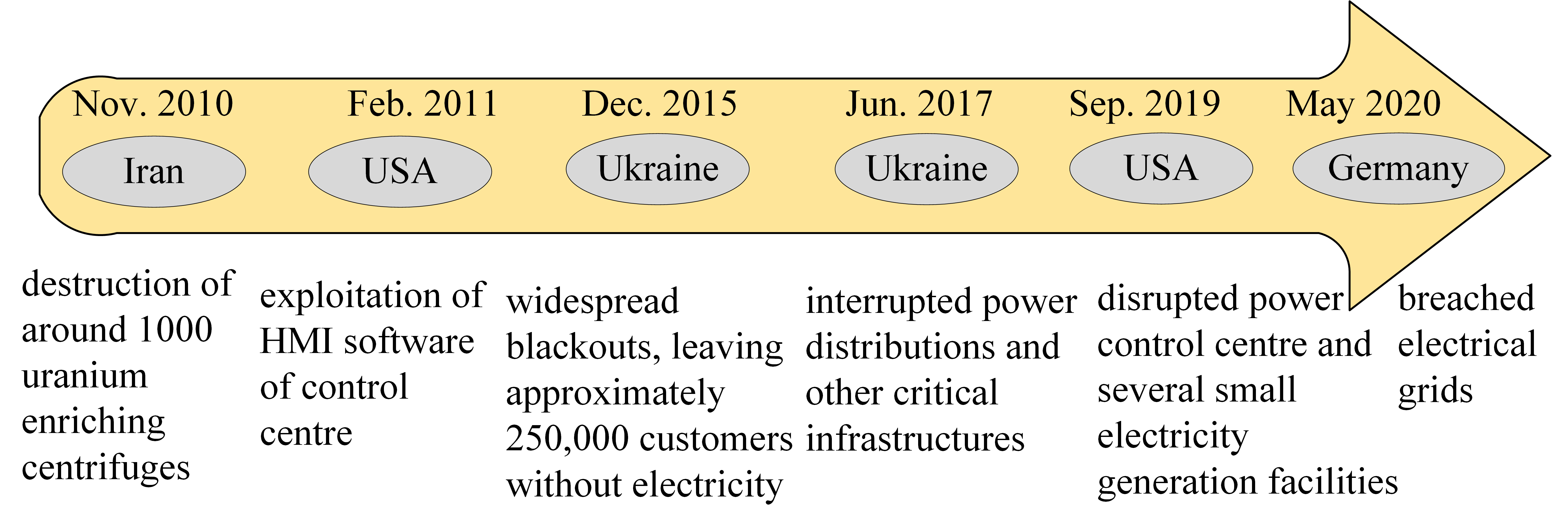}}
	\caption{Timeline of recent global cyberattacks on power grid (referred from \cite{iran2011}, \cite{ukrainAtt2015}, \cite{ukrainAtt2017}, and \cite{cyberattck2020}}
	\label{globalAttks}
\end{figure}

There have been various attacks against the power grid over the last decade. Fig. \ref{globalAttks} demonstrates the timeline of the recent global cyber incidents. 
\subsection{Purpose and Scope of the Study}
Bad data detection (BDD) \cite{6655273} \cite{7868276} \cite{7232283} has been widely utilized in the power system control centers for the identification of cyber anomalies. Nevertheless, it has been proven that the BDDs are incapable of detecting false data injection (FDI) \cite{liu2011false} attacks. The primary objective of this article is to provide a systematic literature review and insights into taxonomies of various FDI attack construction methodologies, attack target, and attack impact relevant to the area of Smart Grid cybersecurity.
\subsection{Contributions}
This report has analyzed a great number of publications and reference materials in the attack construction, targets, and impacts of the false data attacks across various domains of the Smart Grid infrastructure. We systematically search for older and more recent related literature, analyse the main findings covered in each literature, critically evaluate them, and compare each solution within the broader conception of the cyber-physical data integrity attacks. Specifically, major contributions of this article are summarised below. 
\begin{enumerate}
\item The paper identifies essential cybersecurity requirements of Smart Grid (Section \ref{CPSattacks}), including theoretical analysis with illustrative examples of stealthy FDI attacks, and requirements for the stealthy FDI attacks (Section \ref{fdiAtt1}). 
\item Following a comprehensive review of relevant existing survey papers, this work highlights their contribution and identifies the gaps that have been addressed through this survey. Detailed comparisons have been highlighted in Table \ref{relW} and the related discussions have been presented in Section \ref{taxoAll}, \ref{attTaerget}, and \ref{attImpct}.   
\item This paper presents three different taxonomies related to the FDI attack models (Section \ref{taxoAll}), attack targets (Section \ref{attTaerget}), and their impacts (Section \ref{attImpct}).
\item This paper analyses the various FDI-based adversarial modeling methods and provides statistical
 facts on the basis of the evaluation criteria in Section \ref{comparisonSec} and Table \ref{criteriaTable}. Furthermore, the paper discusses main research gaps in the existing false data attack papers in Section \ref{comparisonSec}.
\item Finally, this paper provides technical recommendations for emerging advanced application areas, including Internet of Things (IoT)-based Advanced Metering Infrastructure (AMI), cognitive radio, lightweight machine learning (ML) for resource-constrained IoT devices, FDI attack in edge computing environment, impact of FDI attack in distributed electricity trading and Blockchain ecosystem.
\end{enumerate}

We believe that a systematic survey and synthesis of such a large number of independently developed studies will make a major contribution to the Smart Grid cybersecurity discipline.
\subsection{Outline of the Paper}
First, Section \ref{relWor} discusses related survey papers on FDI attacks and compares with our paper. Next, background on Smart Grid and key cyber-physical elements are discussed in Section \ref{backG}. Then, cyber-physical attacks, cybersceurity main objectives, and security requirements of Smart Grid are highlighted in Section \ref{CPSattacks}. In Section \ref{fdiAtt1}, we comprehensively discuss the FDI attack, the attack vector construction methodologies, demonstrate with example the stealthiness of this class of cyber-physical attack, and the  main requirements for the FDI attack under the Smart Grid environment. The next three sections discuss the suggested taxonomy of the FDI attack, mainly from the adversarial point of view. In particular, Section \ref{taxoAll} covers the attack construction model, Section \ref{attTaerget} explores the FDI attack targets, and Section \ref{attImpct} examines the attack impact. Literature search methodology, selection \& analysis of the surveyed literature, and evaluation criteria among the multitude of algorithms of selected surveyed papers are presented in Section \ref{revMethod}. Furthermore, Section \ref{comparisonSec} compares and contrasts among the numerous attack strategies. Following a critical review of the shortcomings found in the literature in Section \ref{litGap}, our technical recommendations that can substantiate future researches in the field are provided in Section \ref{futurePros}. Finally, Section \ref{cncln} concludes this survey article. %Table \ref{notations} lists the abbreviations used in this article.
\begin{comment}
\begin{table}[ht]
\caption{List of abbreviations} 
\centering
\begin{tabular}{m{2 cm} m{5 cm} m{2 cm} m{5.5 cm}}
\hline 
Abbreviation &  Description & Abbreviation &  Description\\ \hline 
AGC & automatic generation control & ML & machine learning \\
AMI & advanced metering infrastructure & MMS & market management system  \\
AWGN & additive white Gaussian noise & NCS  & networked control system  \\
BDD & bad data detection & NIST & national institute of standards and technology  \\ 
CA & contingency analysis & OPF & optimal power flow \\
DEM & distribution energy management & OT & operational technology \\
DER & distributed energy resource & PCA & principal component analysis \\
EMS & energy management system & PMU & phasor measurement unit \\
FDI & false data injection & RTU & remote terminal unit \\ 
GT & grid topology & SAS & substation automation system \\
ICA & independent component analysis & SCADA & supervisory control and data acquisition \\
ICS & industrial control system & SCED & security-constrained economic dispatch \\
IED & intelligent electronic device & SCOPF & security-constrained OPF \\
LMP & locational marginal price & SE & state estimation\\
LR  & load redistribution & WAMS & wide area monitoring system \\
MEP & multi-step electricity price & WSN & wireless sensor network \\
MILP & mixed-integer linear programming  & WLS & weighted least squares\\ \hline 
\end{tabular}
\label{notations} 
\end{table}
\end{comment}
\section{Related Survey Papers}\label{relWor}
The work by D Wang et. al \cite{6837157} is one of the earliest works where authors present a review on the cyber-physical attacks. Authors described the fundamentals of false data attacks from cyber-and physical-side, with cyberattack illustrations being presented on smart meters. Authors in \cite{10.2015.066756} presented a comprehensive survey of FDI attacks under both AC and DC power flow models in Smart Grid. Unlike to previous studies, \cite{10.2015.066756} has overviewed detection schemes and presented on the basis of centralised-and distributed-based state estimation (SE) techniques. Furthermore, a survey research of the data injection attacks with respect to three major cybersecurity aspects, namely FDI attack construction, impacts of the attacks, and countermeasures is studied by R Deng et. al \cite{7579185}. Unlike to previous studies, \cite{7579185} thoroughly studied the impacts of data injection attacks on the electricity market. Another line of survey research is studied in \cite{LIU201735}, which summarises related literature on different attack models, economic impact of the attack, and mitigation techniques for various Smart Grid domains including transmission, distribution, and microgrid networks. Moreover, G Liang et. al \cite{7438916} complement previous studies and discuss various FDI attack models, physical and economic impacts of the attacks, and countermeasures in Smart Grid. Research works in \cite{8744664} and \cite{8766775} also comprehensively discuss the FDI attacks from the attacker's and operator's point of view along with the consequential impacts of the attacks. 

Different from previous surveys the authors of \cite{8887286} reviewed two main classes of detection algorithms: model-based and data-driven, and have discussed the benefits and drawbacks of each technique. As compared to other review works which mostly focus on the energy management system (EMS), the authors in \cite{AOUFI2020102518} discussed FDI attacks on various entities of the online power system security. These authors review and compare studies on the FDI attacks and provide a new class of cyber-oriented countermeasure: prevention (further classified into block chain and cryptography based techniques).
\begin{table*}
\caption{Comparison of current survey articles and our paper}
\label{relW}
\begin{center}
\begin{tabular}{|cc|c|c|c|c|c|c|c|c|c|c|c|} 
\cline{1-12}
& &  \multicolumn{10}{c|}{\textbf{Literature}} \\ \cline{3-12}
\multicolumn{2}{|c|}{\multirow{1}{*}{\textbf{Comparison attributes}}} & \cite{6837157} & \cite{10.2015.066756} & \cite{7579185} & \cite{7438916} & \cite{LIU201735} & \cite{8744664} & \cite{8766775} & \cite{8887286} & \cite{AOUFI2020102518} & \pbox{0.8cm}{Our paper} \\ \cline{1-12}
\multicolumn{1}{|c}{\multirow{7}{*}
{{\textbf{FDI attack model}}}} &
\multicolumn{1}{|c|}{Complete information} & $\checkmark$ & $\checkmark$ & $\checkmark$ & $\checkmark$ & $\checkmark$ & \ding{53} &$\checkmark$ & \ding{53} & $\checkmark$& $\checkmark$ \\ \cline{2-12}
\multicolumn{1}{|c}{} &
\multicolumn{1}{|c|}{Partial information} & \ding{53} & \ding{53} & \ding{53} &$\checkmark$ &$\checkmark$ &\ding{53} &$\checkmark$ &\ding{53} &$\checkmark$ & $\checkmark$ \\ \cline{2-12} &
\multicolumn{1}{|c|}{LR attack} &\ding{53} & \ding{53}&\ding{53} & $\checkmark$&$\checkmark$ &\ding{53} & \ding{53} &\ding{53} & \ddag& $\checkmark$\\ \cline{2-12}
\multicolumn{1}{|c}{} &
\multicolumn{1}{|c|}{GT attack} &\ding{53} &\ding{53} &\ding{53} & \ding{53}&\ddag & \ddag& \ding{53} & \ddag & $\checkmark$& $\checkmark$\\ \cline{2-12}
\multicolumn{1}{|c}{} &
\multicolumn{1}{|c|}{Data-driven}& \ding{53} & \ding{53} & \ding{53} & \ding{53}&\ding{53} & \ddag & \ddag & \ding{53}& \ding{53} & $\checkmark$\\ \cline{2-12} 
\multicolumn{1}{|c}{} &  
\multicolumn{1}{|c|}{Centralised} &\ddag &$\checkmark$ &\ddag &\ddag &\ddag &\ddag & \ddag& \ddag&\ddag & $\checkmark$\\ \cline{2-12}
\multicolumn{1}{|c}{} &
\multicolumn{1}{|c|}{Distributed} & \ding{53} &$\checkmark$ & \ding{53} & \ding{53} & \ddag & \ding{53} & \ding{53} & \ddag & \ddag & $\checkmark$\\ \cline{1-12}
\multicolumn{1}{|c}{\multirow{8}{*}
{{\textbf{FDI attack target}}}} &
\multicolumn{1}{|c|}{EMS} & \ddag & \ddag &\ddag & \ding{53} & \ddag & \ding{53}& \ddag & $\checkmark$ & \ding{53}& $\checkmark$ \\ \cline{2-12} &
\multicolumn{1}{|c|}{AGC} &\ding{53} &\ding{53} &\ding{53} & \ding{53}&\ding{53} &\ding{53} & $\checkmark$& $\checkmark$&$\checkmark$ &  $\checkmark$\\ \cline{2-12} &
\multicolumn{1}{|c|}{DEM} &\ding{53} & \ding{53}& \ding{53}& \ding{53}& \ddag& \ding{53}&\ding{53} &\ding{53} & \ding{53}&  $\checkmark$\\ \cline{2-12} &
\multicolumn{1}{|c|}{MMS} &\ding{53} &\ding{53} & \ding{53}& \ding{53}& \ding{53}& \ding{53}& \ding{53}& \ding{53}& $\checkmark$&  $\checkmark$\\ \cline{2-12} &
\multicolumn{1}{|c|}{{Network comm.}} &\ding{53} &\ding{53} &\ding{53} & \ding{53} & \ding{53}& \ddag&\ding{53} & \ddag&\ddag & $\checkmark$ \\ \cline{2-12} &
\multicolumn{1}{|c|}{Intelligent devices} & \ding{53}& \ding{53} &\ding{53} & \ding{53}&\ding{53} &\ddag &\ding{53} & \ding{53}&\ding{53} & $\checkmark$\\ \cline{2-12}
\multicolumn{1}{|c}{}                        &
\multicolumn{1}{|c|}{Renewable DER} &\ding{53} &\ding{53} &\ding{53} &\ding{53} & \ding{53} & \ding{53} & \ding{53} &\ding{53} &\ddag &$\checkmark$ \\ \cline{1-12}
\multicolumn{1}{|c}{\multirow{5}{*}
{{\textbf{Impact of FDI attack}}}} &
\multicolumn{1}{|c|}{\pbox{1.7cm}{Secure operation\& stability}} &\ding{53} & \ding{53}& \ding{53}& \ding{53}& \ddag& \ddag& \ding{53}&$\checkmark$ &\ddag & $\checkmark$ \\ \cline{2-12} &
\multicolumn{1}{|c|}{Risk and reliability} &\ding{53} & \ding{53} &\ding{53} &\ding{53} &\ddag &\ding{53} & \ding{53}& \ding{53}& \ding{53}& $\checkmark$\\ \cline{2-12} &
\multicolumn{1}{|c|}{Electricity market} &\ding{53} & \ding{53}&$\checkmark$ &\ddag&$\checkmark$ & \ding{53}&\ding{53} &\ding{53} &\ding{53} & $\checkmark$\\ \cline{2-12} &
\multicolumn{1}{|c|}{Energy theft} &\ding{53} & \ding{53}&\ding{53} & \ddag&\ddag &\ding{53} &\ding{53} & \ding{53}&$\checkmark$ &$\checkmark$ \\ \cline{2-12} &
\multicolumn{1}{|c|}{Energy privacy} &\ding{53} & \ding{53}& \ding{53}&\ding{53} & \ding{53}& \ding{53}& \ding{53}& \ding{53}& \ding{53}& $\checkmark$  \\ \cline{1-12}
\multicolumn{2}{|c|}{\textbf{Evaluation criteria}} &
 \ding{53} & \ding{53} & \ding{53} & \ding{53} & \ding{53} & \ding{53} & $\ddag$ & $\ddag$ & $\ddag$ & $\checkmark$ \\ \cline{1-12}
\multicolumn{2}{|c|}{\textbf{Future directions}} & \ding{53} & \ding{53} & \ding{53} & $\checkmark$ & $\checkmark$ & \ding{53} & $\checkmark$ & $\checkmark$ & $\checkmark$ & $\checkmark$ \\ \cline{1-12} 
\multicolumn{2}{|c|}{\textbf{Duration of surveyed papers}}  &
 \pbox{0.5cm}{2009 to 2013} & \pbox{0.5cm}{2009 to 2013} & \pbox{0.5cm}{2009 to 2015} & \pbox{0.5cm}{2009 to 2015} & \pbox{0.5cm}{2009 to 2016} & \pbox{0.5cm}{2010 to 2017} & \pbox{0.5cm}{2010 to 2019} & \pbox{0.5cm}{2011 to 2019} & \pbox{0.5cm}{2009 to 2019} & \pbox{0.5cm}{2009 to 2020} \\ \cline{1-12}
\end{tabular}
\begin{tablenotes}
  \item \: \: \: $\checkmark$ studied/covered,  \ddag \:partially studied, \ding{53} not studied 
  \end{tablenotes}
\end{center}
\end{table*}
Unlike to the related works, this paper presents a detailed survey of recent developments in the FDI and sets out a taxonomy of the incumbent cyberattack with respect to adversarial models, attack targets, and impacts across every Smart Grid domain including transmission to consumption, automatic generation control (AGC) to microgrids or distributed energy resources (DERs), substation to wide area monitoring systems. IoT, cognitive radios, and software-defined networks have recently been introduced as enablers to the Smart Grid. These communication technologies are very important to address the cybersecurity aspects of today's Smart Grid which were missed in most of the existing related works. In general, in light of research, this paper provides an in-depth survey of the latest advances of the cyber-physical FDI attacks within the Smart Grid infrastructure. Table \ref{relW} summarises the comparison of existing survey papers and this article.
\section{Background} \label{backG}
Smart Grid is primarily the convergence of two interdependent layers (i.e. cyber and physical systems), which are bound together and create a cyber-physical ecosystem. It is crucial to scrutinize the relations between the physical and the cyber entities in order to investigate any underlying cyber-physical attack incidents. Therefore, in this section, the main cyber-physical elements of the Smart Grid are briefely discussed.   
\subsection{SCADA}
Supervisory control and data acquisition (SCADA) \cite{101} is an industrial and power system control application. Usually a SCADA consists basically of three subsystems: a data acquisition sub-system that collects measurement of the power system, a supervisory sub-system that can control remote  intelligent electronic devices (IEDs) \cite{101} by transmitting control commands, and a communication sub-system that interconnects the data acquisition sub-system to the supervisory sub-system. A typical scenario in the integrated SCADA system can be described, for example, when the SCADA gathers data from diverse IEDs in a power system through various communication methods, and then monitor the data using different visualisation tools.
\subsection{Energy Management System}
Power system operations are regulated by system operators from the control center. Within the control center lies EMS, an automation system used to monitor, control, coordinate, and optimize energy data performance across the majority of Smart Grid infrastructure in real time. EMS depends on a SCADA system for its data monitoring
and analysis events. A typical EMS comprises the following functional elements including SE, optimal power flow (OPF), contingency analysis (CA), alarm management system, planning and operations, AGC and economic dispatch.

\begin{figure*}[!ht]
	\centerline{\includegraphics[width=160mm, scale=1.2]{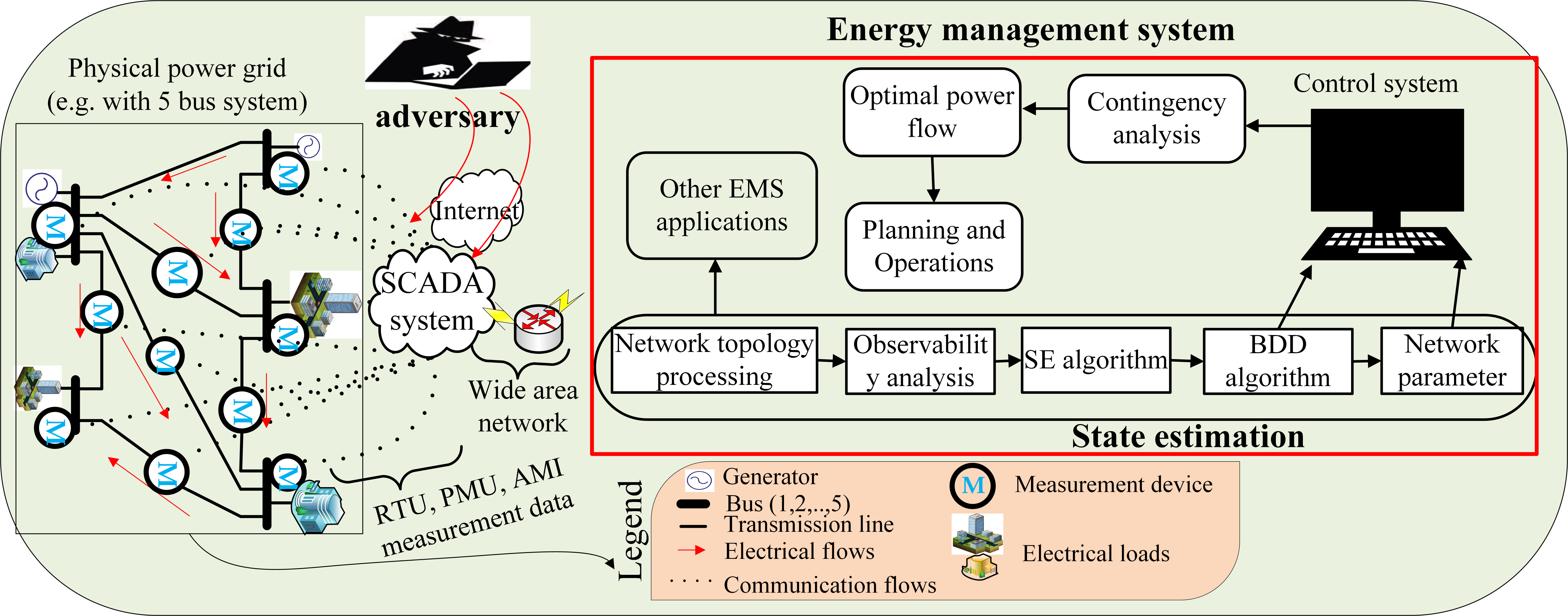}}
	\caption{Typical Smart Grid with 5-bus system.}
	\label{motSE}
\end{figure*}

At the control center, a state estimator receives a set of measurements, and provides accurate system information and detects malicious measurement data (usually through a BDD \cite{6655273} \cite{7868276} \cite{7232283}). In general, the SE outcome presents real-time database for other EMS applications (see Fig. \ref{motSE}).
\subsection{Smart Grid Communication Systems}
Communication systems are essential to the efficient operation of the Smart Grid. Various communication technologies are utilised across the different domains, including IEC 61850 \cite{8757972} \cite{s21041554} in substation automation system (SAS), PMU \cite{wang2019two} in wide area monitoring systems (WAMS), AMIs \cite{6298960} across customer-side, and Networked Control System (NCS) \cite{8766208} between sensors, actuators and controllers.
\subsection{Distributed Energy Resources (DERs)}
DERs are decentralised, versatile, and modular architecture that incorporate a number of renewable sources \cite{6298960}. Compared to conventional approaches in which energy is generated by centralised and big power plants, DER allows energy production and delivery from many areas, including millions of homes and businesses. Microgrid technology is one of the enablers of Smart Grid that provides smooth collaboration between DERs offering isolation options or access to the conventional grid electricity.
\section{Cyber-Physical Security of Smart Grid} \label{CPSattacks}
The security issues of Smart Grid have emerged from both
physical and cyber spaces that include:
 physical security \cite{1213} (i.e. security policies with respect to staffs or personnel, physical equipment protection, and contingency analysis), 
  cybersecurity (focusing on the information security of Smart Grid pertaining to IT, OT, network and communication systems), and
  cyber-physical security (incorporating strength in all physical and cybersecurity measures against inadvertent cyber-physical incidents within an integrated Smart Grid framework). 
In this section, Smart Grid cybersecurity goals, cybersecurity requirements, and cyber-physical attacks are highlighted.
\subsection{Smart Grid Cybersecurity Goals}
Quality of service and secure power supply are the primary concern of power companies and industrial sectors. So much that the Smart Grid strives to build a much more efficient and reliable energy, cybersecurity threats can inevitably slow down its progress. Therefore, the Smart Grid needs to ensure the basic security goals such as data integrity, availability confidentiality, accountability, and etc of the various cyber-physical elements. While these security principles have been developed to govern policies on generic information security within organisations, the principles of Smart Grid cybersecurity have also been identified by NIST \cite{1213}.

\textbf{Avaiablity}:
The permanent availability and timeliness of electricity are crucial in our day to day life. Within the Smart Grid environment, availability is by far the most critical security goal for stability of the power grid. It ensures reliable access to and timely use of information. Availability can be quantified in terms of latency, the time required for data to be transmitted across the power grid. Smart Grid cybersecurity solutions should provide acceptable latency thresholds of various applications by minimising detrimental effects on the availability.

\textbf{Integrity}:
Integrity is the second yet highly critical Smart Grid security requirement. As part of the cybersecurity objectives, integrity ensures that data should not be altered without authorized access, source of data need to be verified, the time stamp 
linked with the data must be identified/validated, and quality of service is under acceptable range.

\textbf{Confidentiality}:
From the point of view of system reliability, confidentiality seems to be the least important as compared to availability and integrity. Nevertheless, with the proliferation of smart meters and AMIs across the Smart Grid implies the increasing importance of confidentiality to prevent unauthorized disclosure of information, and to preserve customer privacy or proprietary information.   

\textbf{Accountability}: 
Another security objective within the Smart Grid ecosystem is accountability, a requirement that consumers should be responsible for the actions they take. Accountability is very important, particularly when customers obtain their billing information from the utility center, they will have sufficient evidence to prove the total power load that they have used.
\subsection{Smart Grid Security Requirements}
The dynamics of the cyber-physical interaction in the Smart Grid poses extrinsic system dependencies. Further, the open inter-connectivity of Smart Gird with the Internet brings various security challenges. Therefore, Smart Grid requires stringent holistic security solutions to uphold the security objectives discussed above and to provide salient features within the Smart Grid infrastructure. First of all, the security solutions need to be robust enough to counteract against increasing security breaches that can lead to loss of data availability, loss of data integrity, loss of data confidentiality. In other words, the operation of power system should continue 24/7 regardless of cyber incident maintaining the power grid reliability (consistent to the data availability and to almost 99.9\% \cite{1213} of data integrity across the power system), and ensure consumer privacy. Second, resilient cyber-physical operations are required. According to NIST's recommendation \cite{calder2018nist}, cybersecurity in critical infrastructure such as the Smart Grid can adopt a comprehensive security framework containing five main features. These include identifying of risks or cyber incidents, providing protective mechanisms against the impact of a potential cybersecurity event, providing defence mechanisms to allow prompt discovery of security breaches, appropriate response to minimise the effect of the incident, and recovery plans to restore any systems that have been disrupted due to cyber accidents. Moreover, as attacks from cyber criminals on the power grid continue to rise in complexity and frequency, it is inevitable that various parts of the Smart Grid are vulnerable to the incumbent attacks. Therefore, it is required to provide strong attack defence across the EMS and to deploy secure communication protocols.  
\subsection{Cyber-Physical Attacks on Smart Grid}
Attacks on Smart Grid vary on a wide range of factors, such as the attacker's motive, capability, skill, and familiarity with the cyber-physical system processes. Different cyber criminals generate attacks based on their ease of attack implementation, course of incidents, and less complexity of creating the attack to maximise damage. Multiple Smart Grid components are likely to be vulnerable to simultaneous cyber threats which could result in widespread power outages. The various types of cyber-physical attacks against Smart Grid are broadly divided into attacks on data availability, attacks on data integrity, attacks on data confidentiality,  accountability and authentication attacks. 
\section{False Data Injection Attacks} \label{fdiAtt1}
FDI attack, first suggested by Liu et al.\cite{liu2011false}, is one of the most critical malicious cyberattacks in the power system. The theoretical frameworks for false data attacks are discussed in this section.  
\subsection{Stealthy FDI Attack} \label{stealthyy}
After SE is conducted, BDD techniques are employed to identify any injected bad data by computing residual vectors in terms of $\ell_2$-norm\footnote{$\ell_2$-norm of \textbf{r} is defined as \begin{math}
 ||\hat{\textbf{r}}||_2^2 = \sqrt{\sum {\hat {\textbf{r}}}^2}
\end{math}} between the original measurements $\mathbf{y}$ and the estimated measurements $\mathbf{\hat{y}=H\hat{x}}$, given by $\mathbf{||r||_2^2=||y-H\hat{x}||_2^2}$. However, research \cite{liu2011false} proved that BDDs are vulnerable to FDI anomalies. The outstanding feature of false data attacks is the residual vectors of the SE drop below the BDD's threshold despite the presence of maliciously corrupted measurements. Consequently, such strategically constructed false data attack vectors can bypass (i.e. remain stealthy in) the traditional BDD algorithms.
\subsubsection{FDI Attack Construction and Proof of Stealthiness}
In the presence of FDI attack, the adversary's goal is to introduce an attack vector \textbf{a} into the measurements without being noticed by the operator. Adversaries approach with different FDI attack strategies whereby the final effect of the malicious data results in compromising state variables across the power system domain. Generally, there are two main FDI anomaly construction strategies, one that requires knowledge of power system topology, and the other is based on a data-driven approach also known as the blind FDI attack strategy (details are given in Section \ref{taxoAll}). Here, we use the former approach to demonstrate the stealthiness of the FDI attack. Let $\textbf{a}=[a_1, a_2,...,a_m]^T$ denotes the FDI attack, then measurements that contain this malicious data are represented by  
\begin{math}
 %\label{malMeas}
 \textbf{y}_{false} = \textbf{y}+\textbf{a},  
\end{math} 
and 
\begin{math}
 %\label{malState}
 \hat{\textbf{x}}_{false} = \hat{\textbf{x}}+\textbf{b}  
\end{math} refers to the estimated state vector after the FDI attack, where \begin{math}
  \textbf{b} = [b_1, b_2, ..., b_n]^T
\end{math} is the estimated error vector injected by the adversary. It is usually assumed \cite{liu2011false} that the attack vector \textbf{a} can be formulated as a linear combination of \textbf{H} given by \textbf{a} = \textbf{H}\textbf{b}. 
It has been proven \cite{liu2011false} that
if \begin{math}
 ||{\textbf{r}}||_2^2 < \tau
\end{math} it also holds true that \begin{math}
 \mathbf{||r_{\text{false}}||_2^2 } < \tau 
\end{math} for some detection threshold $\tau$. Hence, under \textbf{a} = \textbf{H}\textbf{b} the malicious measurement vector can pass the traditional BDD algorithms. 
\begin{figure*}[!ht]
	\centerline{\includegraphics[width=160mm, scale=1.2]{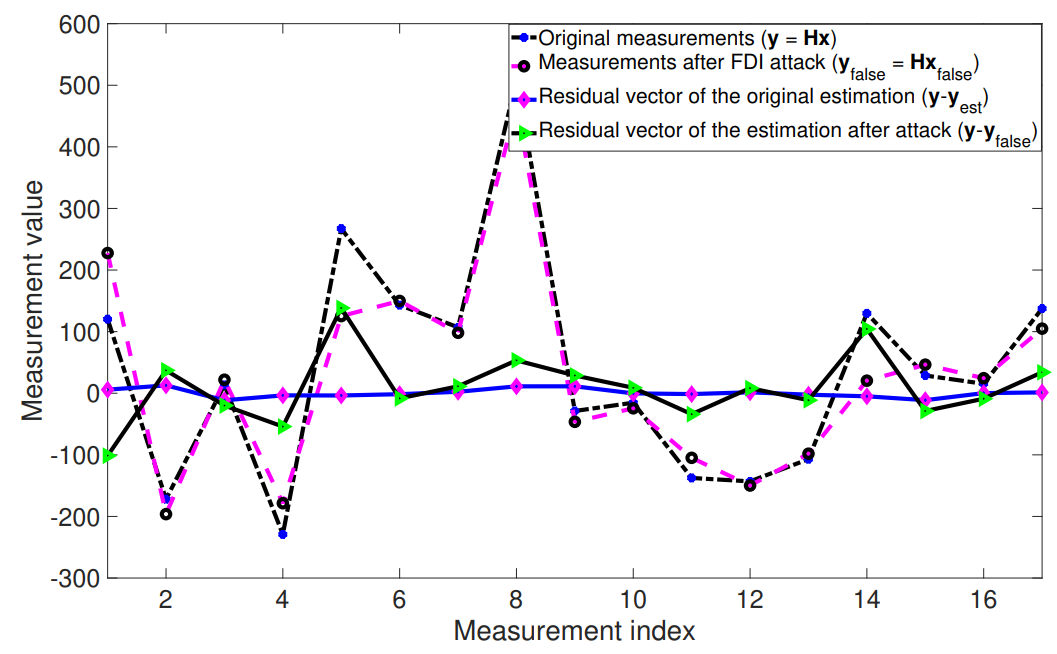}}
	\caption{Power system measurement (before and after FDI attack.}
	\label{fig5}
\end{figure*}
\begin{figure*}[!ht]
	\centerline{\includegraphics[width=160mm, scale=1.2]{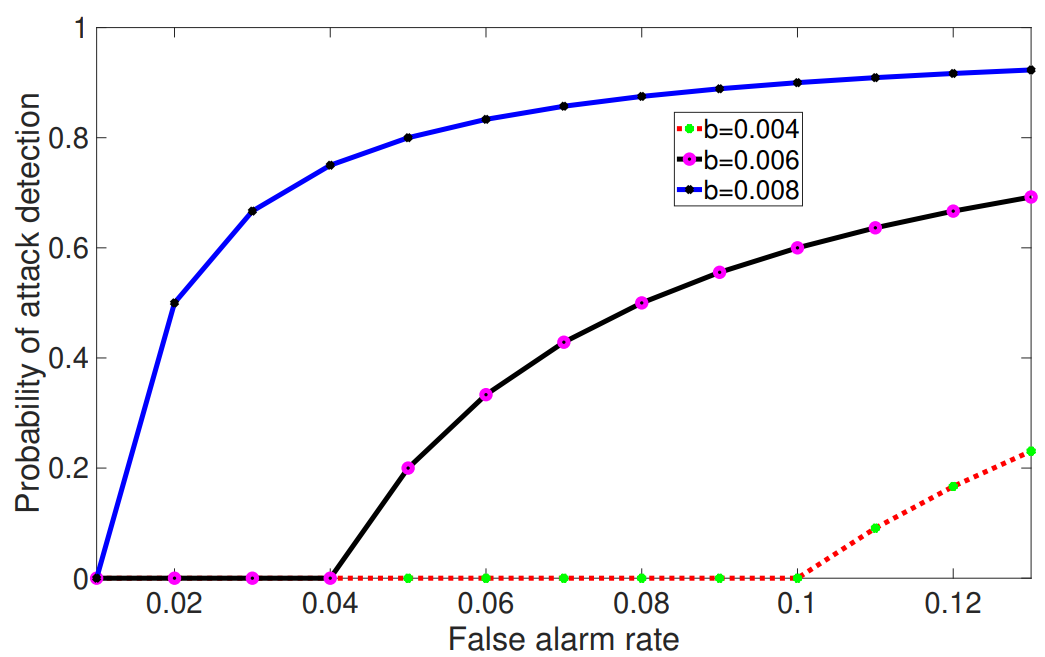}}
	\caption{Attack detection based on $\chi^2-$detector.}
	\label{fig6}
\end{figure*}
\subsubsection{Sparsity of FDI Attack}
Although \textbf{a} is usually assumed as a linear combination of the columns of \textbf{H} \cite{liu2011false}, the adversary's control can be limited to only over a few measurement devices (let us say $k$). It could be because either the system has secure measurement devices which the attacker cannot access, or the attacker has limited physical access to the devices. This results in a sparse FDI attack \cite{liu2011false} \cite{ozay2013sparse}. FDI attack designed with only few non-zero components is called sparse attack. 
\subsubsection{Demonstration with an Example}
Here, the operation of the state estimator and BDD module is demonstrated. The aim of this demonstration is to see the effect of the FDI attack on power system measurement and to justify the theoretical discussions that is presented in the aforementioned sections. In this case, IEEE 5 bus system is used as a test case. Fig \ref{fig5} demonstrates measurement results when passed through a weighted least squares (WLS) \cite{Abur} based state estimator both before and after the introduction of stealthy FDI attack. In Fig. \ref{fig5}, four results are shown, the first being the original measurement, the second is the measurement estimation using WLS estimator considering the FDI attack. Further, two residual vectors accounting for the difference between original measurement and attack-free estimated measurement, and the difference between original measurement and FDI-estimated measurement are shown. The attack free residual result shows approximate to zeros where as the compromised measurements have shown a very large deviation. In addition, in Fig \ref{fig6}, Chi-square ($\chi^2$)$-$detector is employed for the detection of bad data. The detection result shows a very poor probability of detection of the false data attack.
\subsubsection{Observability}
System observability is an important requirement for the operator. The observabiltiy analysis determines whether a unique estimate can be calculated for the system state provided that adequate measurements are available in the power system \cite{Abur}. Therefore, when there are adequate measurement observations to determine the values of the state variables, the power system is said to be fully-observable. In contrast, limited number of measurements are observed due to either removal of maliciously compromised measurements or limited number of PMU placements in the field which results in partial observability or even unobservable system. 

A systematic amalgamation of injected data with the original power system measurement data by cyber adversaries produces a falsified estimate of the critical power states over the control centers. This is because attackers can intelligently construct the FDI attack vectors (as it has been shown above) that can cause the power system unobservabile. 
\subsection{Requirements for Stealthy FDI attacks} \label{reqFDI}
The requirements of FDI attacks vary from one application domain to another. For example, in wireless sensor networks (WSNs), the inherent wireless communication and broadcast channels between the nodes increase the vulnerability of adversaries that may eavesdrop on all traffic, inject false data reports containing erroneous sensor readings, or can even deplete the already limited energy capacity of sensor nodes \cite{1301328WSN}. On the other hand, in the power system, it is challenging for an adversary to access the network parameters, and thus needs a much more intelligent strategy in order to launch a successful attack. In general, therefore, the FDI attacks impose strong requirements both from the point of view of adversaries and the system operators. The following are some of the main requirements for the stealthy FDI attacks in the cyber-physical Smart Grid environment.
\subsubsection{Rendering power system unobservability \cite{liu2011false}} Through the injection of false data, the attacker can remain undetectable at the control center while resulting in incorrect decisions of the state estimator. Even if the cyberattack can be detected by the SE, part of the power network may become unobservable where the SE cannot determine the system states.
\subsubsection{Partial-Parameter-Information} Earlier studies on the FDI attack models are based on the premise that the adversaries are capable of getting complete information of the power system topology. Authors in \cite{6503599DCincompleteInfo1} presented that it is also possible to construct stealthy attacks based on partial network information. Yet, attacks based on partial information need to satisfy the observability criteria. Another research direction ensures that the stealthiness (i.e. undetectability) of FDI attacks can also be modeled through data-driven or other partial-parameter-information approaches.   
\subsubsection{Minimal Attack Vectors} For many reasons, the adversary's control can be limited to only over a few measurement devices. It could be because either the system has secure measurement devices which the attacker cannot access, or the attacker has limited physical access to the devices. For this reason, stealthy FDI attacks should be designed with a very small attack magnitude and with only few non-zero components (i.e. attack sparsity) \cite{liu2011false} \cite{ANWAR201758PCA4}. Consequently, the attacker is required to compromise just smallest set of devices to cause network unobervability. 
\subsubsection{Attack Specificity} 
Whatever the motives of the cyber criminal are, the strategy behind the attack may be either indiscriminate  or targeted. The scope and impact of these two adversarial approaches are different. For example, in the former, the FDI attack may not require specific knowledge of the cyber-physical devices but launched arbitrarily against random Smart Grid elements. On the other hand, the targeted one can require a sophisticated approach which can be launched against targeted nodes or communication infrastructure or any targeted cyber-physical element. One of the most prominent targeted FDI attacks is load redistribution attack \cite{yuan2011modeling} targeting load measurements of nodal power injections and power flows.  
\subsubsection{Requirement on The Influence of The Attack} Attackers can approach in various ways to launch a successful attack and to cause a security risk on the Smart Grid. Some attackers want to exploit the data collected from sensors and networked devices across the power system. They may intend to exploit the weaknesses of sensors and communication protocols and launch the attack vector. Some typical examples of attack scenarios can be attack against sensor measurements (tampering power system parameter values in remote terminal units (RTUs) and PMUs). Another example can be by leveraging the communication protocols, where remote tripping injection can be performed by adversaries. In addition, attackers can infiltrate AMI-based communications networks in order to tamper with the contents of customer data that can result in disorder of the SE and other EMS functionalities. Others may intend to directly falsify the outcome of the state estimators \cite{liu2011false}.
\subsubsection{Requirement Based on Security Violations}
Some FDI-based malicious attackers try to infringe data availability, some violate data integrity, and others go against data confidentiality.  
\begin{enumerate} [label=(\alph*)]
\item Loss of data integrity:  For example, by injecting a systematically generated false data, a cyber intruder may compromise the integrity of the SE by hijacking a subset of metres and returning a modified data. The modification may involve deletion of data from the original meter readings, addition of bad data to sensor readings, or alteration of values in the hijacked measurements. The majority of FDI attacks, including, but not limited to, \cite{liu2011false} \cite{ANWAR201758PCA4} \cite{GIANI2014155} \cite{7885045} are based on this type of security violation.    

\item Loss of data availability: Furthermore, FDI attack can compromise the availability of critical information that is either intended to disrupt the power system or to stop its availability by shutting down network and communication devices  \cite{10.2015.066756}  \cite{7579185}
\item Attack on confidentiality: Although the effect of FDI on data confidentiality ranks among the least of all security objectives, the injection of false data could also violate the privacy of customers, especially in AMIs of the Smart Grid. This has become so common these days as illustrated in \cite{8918446}  \cite{chen2020blockchain}.
\end{enumerate}
\begin{figure*}[!ht]
	\centerline{\includegraphics[width=145mm, scale=0.2]{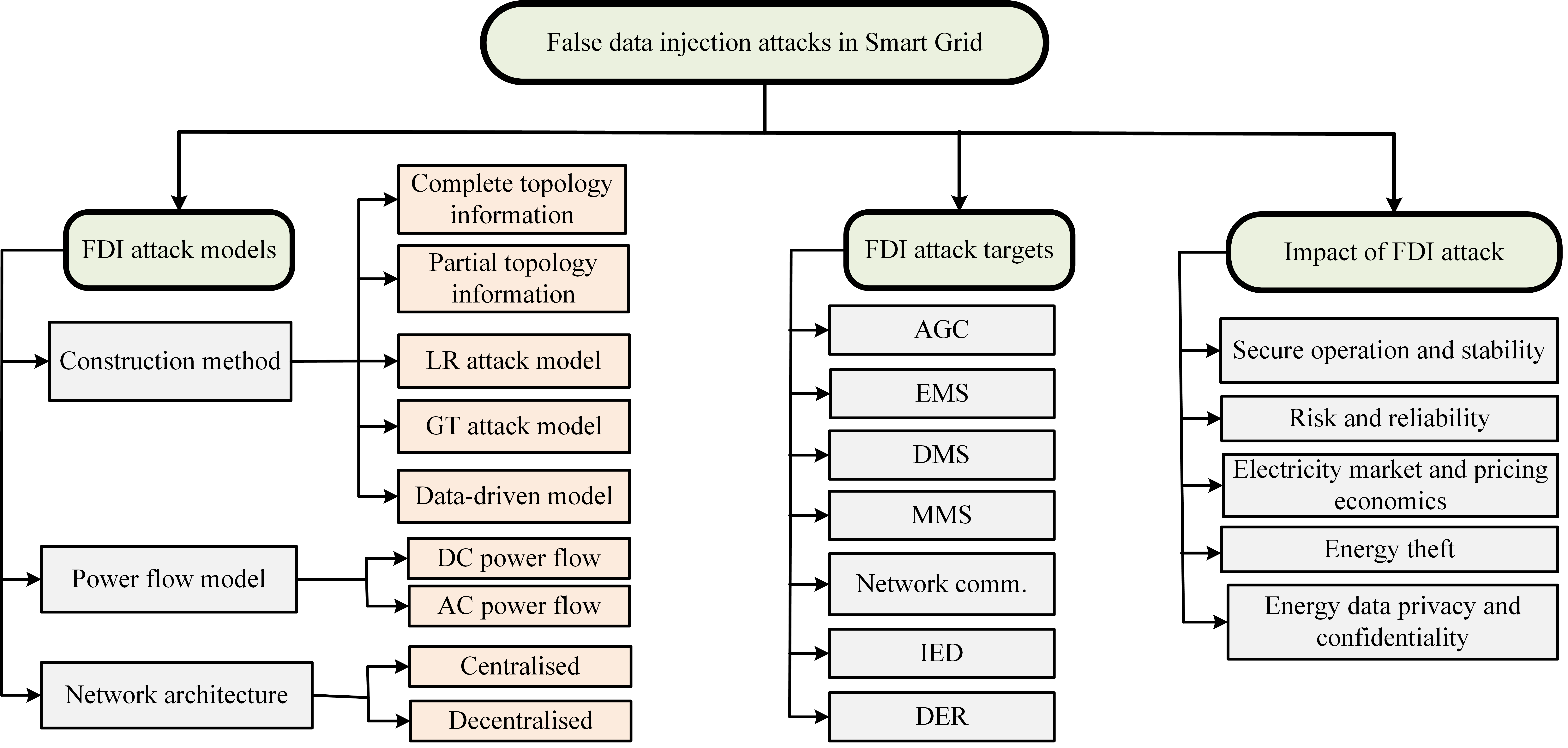}}
	\caption{Taxonomy of false data attacks in Smart Grid}
	\label{taxAll}
\end{figure*}
\subsubsection{Requirement based on Attack Impact on The Power System} 
Threat actors can exploit Smart Grid security vulnerabilities that may lead to malfunctions in energy systems, operational failures in communications equipment as well as physical devices, and may even trigger a cascading failure. According to a report by NIST \cite{1213}, three potential impact levels, namely low, moderate, and high have been assessed for each of the Smart Grid security objectives following the degree of adversarial effect and associated risk level. 

Eventually, the ultimate aim of FDI adversarial strategies is to pose significant consequences against the Smart Grid, such as causing sequential transmission line outages, maximizing operation cost of the system by injecting falsified vectors into subset of targeted meters, culminating in large-scale failure of the power system operation, and regional/national catastrophic impacts.
\section{Taxonomy}\label{taxoAll}
The success of cyber-physical attacks in general and the FDI attack in particular depends on both the perspective of the adversary and the operator. In other words, it is highly likely that adversaries are subject to a trade-off between maximizing the probability of impact on various cyber-physical system components and minimizing the probability of detection of the launched attack. In this survey paper, the false data attacks in Smart Grid are classified under three major categories. 
\begin{enumerate}
\item False data attack models: There are various threat models of FDI to corrupt the cyber-physical infrastructure of the Smart Grid. While some adversarial models require complete information on network data and topological configurations, others require limited resources. Data-driven approaches are also employed to construct the stealthy FDI attacks. This FDI category is presented in this section (\ref{attackModel}). 
\item False data attack targets: Coordinated cyberattackers try to target various elements of the Smart Grid. The vulnerable components include, but not limited to, power generators, transmission lines, substation networks, renewable energy sources, monitoring and control centers, smart electronic devices, network and communication systems, which are all discussed in Section \ref{attTaerget}. 
\item Impact of false data attacks: The growing threat of cyberattacks against the critical Smart Grid infrastructure have devastating impacts on its stability, reliability, economy, customer data privacy, and social welfare. The category of attack impact is addressed in Section \ref{attImpct}.  
\end{enumerate}
Each class is also divided into subcategories. Fig. \ref{taxAll} is the taxonomy of the FDI attacks. The order of presentation of each category is to a large extent a reflection of the chronological order of the researches, starting from attack construction, going through the targets, and attack impacts. While the first taxonomy is discussed in this section, the other two are presented in the subsequent two sections.     
\subsection{Classification Based on Attack Model}
 \label{attackModel}
Adversaries follow various FDI attack strategies whereby the ultimate outcome of their malicious activity results in breaching state variables across the power system domain. The various FDI attack models can be seen in the first part of Fig. \ref{taxAll}. In the following, each false data attack category is described, and Table \ref{table_attModel} lists several of the relevant papers that focus on FDI attack construction methodologies.
\subsubsection{Power Flow Model}
Most FDI attack researches are conducted in a constrained environment, on the basis that the functions from power system states to measurements are linear (DC-based power flow models) while most industry standard state estimators are based on the nonlinear AC power flow model. One of the pioneer FDI attack under the DC model is proposed by Liu et al.\cite{liu2011false}. Since then other similar lines of researches have been studied including  \cite{6503599DCincompleteInfo1} \cite{7741557PCA3} \cite{7387159DCincomleteInfo2} \cite{7031948DCincomleteInfo3} \cite{ANWAR201758PCA4}.   
In most situations, the study of the AC power flow models has to be accompanied by solving complete nonlinear power flow equations which are involved in the nonlinear models. Consequently, the complexity of analysis must be reduced and completely ignored the nonlinear constraints while modeling the cyberattacks. While most of the FDI techniques available in the literature rely on the simplified DC state estimators, such techniques are not valid to AC-based SEs. For example, Hug and Andrew \cite{6275516ACattack} have shown that the nonlinear representation of power systems in AC state estimators have inherent strengths and are more robust to unobservable FDI attacks than the DC-based SEs. They have analytically derived stealthy FDI attack for AC SE model considering RTUs as attack target using IEEE 57 bus test system. Accordingly, adversaries using a specific type of FDI attack like in the RTU level under DC model has higher risk of introducing errors in the measurements, which in effect, would trigger the BDD, and the adversary requires significantly more system data under the AC model than the DC model of the same target. Similarly, authors in \cite{6672638AC1} \cite{7401110ACincompleteInfo1} studied the construction of stealthy FDI attacks against AC-based SE models. Further, in \cite{6939486} the authors have proven that DC-based attacks can be detected using AC-based SE even though the attack magnitude is relatively small. Moreover, it has been investigated that AC-based SE models need a more sophisticated attacker than the DC-based models \cite{7366616AC}. Very recently, the authors of  \cite{doi:10.1002/cpe.5956AC2} studied an experimental case  with industrial standard AC-based SE is utilised to exemplify an AC model based FDI attack design. And, only few researches \cite{9026314PCA2} \cite{7001709Data-drivenPCA1} have studied the FDI attack under both the DC and AC power flow models.      
Therefore, it is highly important that the vulnerability analysis of the Smart Grid with respect to the incumbent cyberattacks, and the countermeasures requires a thorough understanding of the physical properties of the power system in general and which power flow model is utilised.
\begin{table*}[ht]
\caption{FDI attack construction methodologies}
\centering
\begin{tabular}{M{4cm} M{1cm} M{1cm} M{1.2cm} M{1.4cm}M{6cm}}
  \hline \Xhline{2\arrayrulewidth} & \multicolumn{2}{c}{\textbf{Power flow}} &  
  \multicolumn{2}{c}{\textbf{Architecture}}
  \\
  \cmidrule(lr){2-3} \cmidrule(lr){4-5}
\textbf{FDI attack model} & \textbf{DC} & \textbf{AC} & \textbf{Centralised} & \textbf{Decentralised} & \textbf{References}  \\  \hline
  Complete Topology Information & 
  $\checkmark$ & {\ding{53}} & $\checkmark$ & {\ding{53}} & \cite{liu2011false} \cite{8281015} \cite{jiongcong2016impact} \cite{sandberg2010security} \cite{5464816Centralised4} \cite{6032057Centralised2} \cite{5751206Centralised3}  \cite{tan2017modelingAGC4} \cite{6547837topologyAtt1} \cite{khanna2017bi} \cite{basumallik2020impact}  \cite{6074981} \cite{tan2015integrity} \cite{5947717} \cite{6831164} \cite{8219710} \cite{8023151} \cite{5622048} \\ \cline{2-6}
   & {\ding{53}} & $\checkmark$ & $\checkmark$ & {\ding{53}} & \cite{6672638AC1} \cite{7366616AC} \cite{8403288AC} \cite{9285056}  \\ \cline{2-6}
 & {\ding{53}} & $\checkmark$ & {\ding{53}}  & $\checkmark$ & \cite{6655273} \cite{8425789ACincompleteInfo2} \cite{choeum2019oltc} \cite{wang2014extended}  \\  \cline{2-6}
    & $-$ & $-$ & {\ding{53}}  & $\checkmark$ & \cite{nikmehr2019game} \cite{7249479} \cite{7570183}
\\  \cline{2-6} 
  & $\checkmark$ & {\ding{53}} & $\checkmark$ & $\checkmark$ & \cite{ozay2013sparse} \cite{6486001Distributed4} \\ \cline{2-6}
& $\checkmark$ & {\ding{53}} & {\ding{53}}  & $\checkmark$ & \cite{6102319Distributed3} \cite{8625609} \\ \cline{2-6}
  & $\checkmark$ & $\checkmark$ & $\checkmark$ & {\ding{53}} & \cite{6275516ACattack}  \cite{6939486} \\  \cline{2-6}
 & $-$ & $-$ & $\checkmark$ & {\ding{53}} & \cite{8270717} \\  \hline 
Partial Topology Information & 
  $\checkmark$ & {\ding{53}} & $\checkmark$ & {\ding{53}} & \cite{6547837topologyAtt1} \cite{8581440ACincompleteInfo3} \cite{8323244} \cite{7995087} \cite{7446354}  \\   \cline{2-6}
  & $\checkmark$ & {\ding{53}} & {\ding{53}}  & $\checkmark$ & \cite{7031948DCincomleteInfo3} \cite{7438904} \cite{7185427topologyAtt2} \cite{7433442topologyAtt3}  \cite{7271100topologyAtt5} \\  \cline{2-6}
  & {\ding{53}} & $\checkmark$ & $\checkmark$ & {\ding{53}} & \cite{wang2019two} \cite{6672638AC1} \cite{7401110ACincompleteInfo1} \cite{8260948}  \\ \cline{2-6} 
& $\checkmark$ & $\checkmark$ & $\checkmark$ & {\ding{53}} & \cite{7741928}  \\   \hline
LR attack  & 
$\checkmark$ & {\ding{53}} &$\checkmark$& {\ding{53}} & \cite{yuan2011modeling} \cite{6831166} \cite{7286402} \cite{6148224} \cite{6805238} \cite{8468098} \cite{xiang2017game} \cite{lee2019vulnerability} \cite{8338154} \\  \cline{2-6}
  & {\ding{53}} & $\checkmark$ & {\ding{53}} & $\checkmark$ &  \cite{8260848} \\  \cline{2-6}
  & $-$ & $-$ & $\checkmark$ & {\ding{53}} &  \cite{7470429}
\\  \cline{2-6} & $\checkmark$ & $\checkmark$ & {\ding{53}} &  $\checkmark$& \cite{6848210} \\ \hline 
GT attack  
  & $\checkmark$ & {\ding{53}} & {\ding{53}}  & $\checkmark$ & \cite{7438904} \cite{7185427topologyAtt2} \cite{7433442topologyAtt3} \cite{7271100topologyAtt5} \cite{7426376topologyAtt4} \\  \cline{2-6} 
& {\ding{53}} & $\checkmark$ & $\checkmark$  & {\ding{53}} & \cite{7450168} \cite{chung2018local}  \\ \cline{2-6} 
&  $\checkmark$ & {\ding{53}} & $\checkmark$  & {\ding{53}} & \cite{6547837topologyAtt1}   \\ \hline
Data-driven  & 
$\checkmark$ & {\ding{53}} & $\checkmark$ & {\ding{53}} &  \cite{ANWAR201758PCA4} \cite{7741557PCA3} \cite{7001709Data-drivenPCA1} \cite{7307141} \cite{7885051} \cite{6996007subspace}     \cite{10.1007/978-3-319-31863-9_13} \cite{tian2018data} \cite{lakshminarayana2020data}\\   \cline{2-6}
& $\checkmark$ & {\ding{53}} & {\ding{53}}  & $\checkmark$ & \cite{7885051} \\  \cline{2-6}
& $\checkmark$  & $\checkmark$ & $\checkmark$ & {\ding{53}} & \cite{7934033} \\  \cline{2-6}
&  {\ding{53}} & $\checkmark$ & $\checkmark$ &  {\ding{53}}  & \cite{8307441Distributed1} \cite{9094321} \\ \cline{2-6}
& {\ding{53}} & $\checkmark$ & {\ding{53}} & $\checkmark$ & \cite{8248780} \\ \hline \Xhline{2\arrayrulewidth}
\end{tabular}
\label{table_attModel}
 \end{table*}
\subsubsection{Network Architecture}
In general, the operation of a Smart Grid depends upon the availability of information from hierarchically distributed cyber-physical elements and the outcomes of the central control center. It is important to investigate that the FDI attacks from the view point of network architecture: centralised and distributed.

Centralised FDI attacks target against the centralised state estimator. Once the adversary manipulates the measurement reports sent from different communication devices to the control center, the SE fails to estimate the optimal system states which further affects other functional elements such as optimal power flow, economic dispatch, and CA that rely on the SE outcome. A great many of FDI attack construction methodologies are introduced using centralized network architecture, some of which include \cite{liu2011false}, \cite{6032057Centralised2}, \cite{5751206Centralised3}, \cite{7307141} (for a list of them, see Table \ref{table_attModel}).  

However, the centralised attacks may be difficult to be implemented in distribution systems, which require knowledge of local states \cite{8307441Distributed1}. Adversaries may also intend to forge the injection of bad data against the energy system at the supply-side, against energy control commands, and the communication link of energy transmission, and distributed energy routing processes \cite{6197400Distributed2}. Some of the FDI research papers which devote on the distributed architecture are \cite{6102319Distributed3}, \cite{6486001Distributed4}, \cite{6197400Distributed2}, and \cite{ozay2013sparse} (list of others can be inferred from Table \ref{table_attModel}).  
\subsubsection{Construction Methods} \label{constMethod}
Here, various adversarial construction methods are discussed.

\textbf{Attacks with Complete Topology Information}: In these types of FDI attacks, adversaries typically require a complete knowledge of network topology, transmission system parameters, details of SE algorithm, and/or BDD methods. This case presumes the adversary has access to several resources of the electric power system and can successfully construct the FDI attack vector. Although most FDI attack researches consider this type of strategy, it is impractical to assume that adversarial models have access to a large number of measurements. Liu et al. \cite{liu2011false} have demonstrated the constraints faced by adversaries. Accordingly, the adversary can be restrained only to certain set of sensor readings; due to the sensors may have specific physical defences or the adversary may have limited budget to compromise the sensors.

According to the findings of \cite{liu2011false}, the objective of the adversaries may be to randomly inject bad data, where they aim to locate any attack vector so long as it can bring a wrong SE performance of state variables, or to launch more targeted attack vectors, where the adversaries aim to build bad data injections into some chosen state variables. Studies include random and targeted FDI attacks from the SE to other cyber-physical components. In \cite{6197400Distributed2} 
random bad data were injected to distributed system to compromise the supply-demand of energy system. Targeted attacks are discussed in more details later.

Differently, Kosut et al. in \cite{5464816Centralised4} and \cite{6032057Centralised2} view the nature of stealthy FDI attacks as a matter of basic constraint on the detectability of malicious data attacks. Unlike to \cite{liu2011false}, Kosut et al. came up with the concept of a detectability heuristic to find the attacks that would render BDD the most vulnerable provided a specific set of meters controlled by the attacker. An extension to \cite{5464816Centralised4}, they proposed FDI attack algorithm \cite{6032057Centralised2} based on minimal energy leakage by considering two forms of attacks: the strong attack and the weak attack. In the strong attack regime, the adversary compromises a sufficient number of meters such that the system state becomes unobservable by the SE utilising a graph theoretic approach, where as in the weak attack regime, the adversary controls just a limited number of meters.    
However, the FDI attacks pose several stringent requirements against the intruders. For instance, the topology settings of the power system are typically only available at the operator's EMS, whose physical access is strongly restricted and secured. Further, these settings do change very often due to routine normal maintenance of electrical power grid devices and unplanned incidents such as unexpected field device failure. In general, intruders have restricted physical access to most power grid infrastructure and they barely have real-time knowledge with respect to topology configurations and physical states like the transformer tap changes, circuit breakers, and switches. Therefore, attackers need to pursue alternative approaches, which are discussed below.

\textbf{Attacks with Partial Topology Information}: As discussed earlier, the construction of valid FDI attack is subject to certain constraints. Although it is ideally fair to implicitly presume that the topology information can be accessible to the adversary in order to build the attack vector; however, it is more realistic to believe that the adversary has incomplete topology knowledge for certain transmission line networks due to the adversary’s lack of real-time knowledge with respect to topology configurations and physical status like the transformer tap changes, circuit breakers, and switches. Therefore, a realistic FDI attack can be launched with incomplete information as the adversary can have only access to limited resources. Rahman et al. \cite{6503599DCincompleteInfo1} proposed FDI attacks using incomplete knowledge of network topology from both the adversary's and defence point of views. Similar line of researches have been studied including, but not limited to  \cite{7387159DCincomleteInfo2} \cite{7031948DCincomleteInfo3}  \cite{7401110ACincompleteInfo1} \cite{8581440ACincompleteInfo3}  \cite{8425789ACincompleteInfo2} 
 
G. Liang et. al. \cite{7438916} have reviewed various scenarios under which adversaries can get partial topology information necessary to launch a successful FDI of this attack category. One is a manual or online mode \cite{6503599DCincompleteInfo1} where before generating the FDI attack the adversary collects grid topology information either manually or through online where the adversary can use his/her own meters to access the grid. The other is through a market database (extracting the topology information from locational marginal prices). Finally, extraction of \textbf{H} from power flow measurements.  

\textbf{Load Redistribution Attacks}: Under restricted access to specific metres, load redistribution (LR) attack is one of special type of FDI attacks targeting load measurements of nodal power injections and power flows. This kind of FDI attack aims to generate biased load estimates. Yuan et al \cite{yuan2011modeling} are the first to formulate the LR cyberattacks with various attacking resource limitations. This framework was further developed in reference \cite{6148224} of the same authors to quantitatively evaluate two attacking goals: immediate and delayed attacks, using a a max-min attacker-defender model. In addition, Xiang et al \cite{7286402} suggested a coordinated cyber-physical attack on LR, generator, and transmission line, formulated as a bilevel optimization problem of attacker-defender model. Also, in reference \cite{6805238}, by using their proposed local topology attacks in \cite{7426376topologyAtt4} and applying the idea in \cite{6503599DCincompleteInfo1}, the authors came up with a local LR attacking strategy with partial network knowledge. Unlike \cite{6503599DCincompleteInfo1}, the attacking region is no longer limited within a cut, for the attacker can select an attacking area of their interests.

\textbf{Grid Topology Attacks and Line Outages}: Attacks against power grid topology (GT) and outages of transmission line are very recent research developments. Most of the adversarial models mentioned earlier are focused on the premise that the  power grid topology stays unchanged. This implies that the adversary can only inject false data to the measurement data of the power system. As a matter of fact, topology configurations do change very often due to routine normal maintenance of electrical power grid devices and unplanned incidents such as unexpected field device failure. Therefore, the state of art literature on FDI attack strategy targeting power system states has further been extended to reflect on the real-time grid topology. The purpose of such attack is to concurrently alter the measurements of network and the topology configurations such as physical states of transformer tap changes, circuit breakers, and switches so that the estimated topology is consistent with the received network data. Such stealthy malicious attack model was formulated by J. Kim and L. Tong \cite{6547837topologyAtt1}. Their proposed adversarial model is characterised by two attack regimes: strong and weak attacks, depending on the information available to the attacker. To avoid the detection by the SE, FDI attack is constructed to make the received measurement data are consistent with the topology while actually aiming to create a false topology at the state estimator under DC and AC power flow models.   

Following the research in \cite{6547837topologyAtt1}, few similar researches, but with a different approach have been conducted. In \cite{7185427topologyAtt2}, the authors studied a coordinated cyber-physical attack that could cause undetectable transmission line outages. They have shown that an adversary can hide the topology of a power grid by injecting bad data into a specific number of measurements. After physical attacks are launched, cyberattacks consisting of topology preserving attack and LR attack are systematically orchestrated to hide line outages and to potentially cause cascading failures. The works in \cite{6547837topologyAtt1} and \cite{7185427topologyAtt2} lack a realistic topology attack to simulate the attack behaviors and to further determine how much network knowledge is required by the adversary to initiate the topology attack. To alleviate this drawback, a topology attack model was suggested in \cite{7426376topologyAtt4}. \cite{7426376topologyAtt4} proposed a heuristic method for determining the possible attacking region of a line using less information. Yet, the researches in \cite{6547837topologyAtt1}, \cite{7185427topologyAtt2}, and \cite{7426376topologyAtt4} did not consider the function of PMUs in detecting the line outages. For example, if a line outage occurs, there would be a deviation in the PMU bus phasors that helps the operator to detect the line outage. Following the same principle as in \cite{7185427topologyAtt2} and \cite{7426376topologyAtt4}, the authors of \cite{7433442topologyAtt3} came up with a concept that line outages can be masked through manipulating PMU data based outage detection by injecting malicious data into  measurements. Additionally, \cite{7271100topologyAtt5} considered the effect of security constrained economic dispatch on the transmission line
 attacking strategy.
 
\textbf{Data-Driven Attack}: In this type of class attack, also known as the blind attack method, undetectable FDI attacks are constructed without prior power grid knowledge, typically using statistical inferences (e.g independent component analysis \cite{hyvarinen2001independent}, subspace-based singular value decompistion \cite{80966}, principal component analysis techniques \cite{abdi2010principal}, sparse optimization \cite{6740901SparseOpt1}),  heuristic methods, and ML algorithms. In other words, the adversary is expected to make inferences from the correlations of measurement data and/or topology parameters of the power system. The question, therefore, is if \textbf{H} is not completely or partially available to adversaries, how can the adversaries still effectively launch the undetectable FDI attack? 

Esmalifalak et al \cite{7307141} are the pioneers to answer the above question. They proposed an inference algorithm using independent component analysis under very small power system dynamics and a linear measurement model. Their findings have shown that an adversary can infer both system topology and power states just by observing the power flow measurements. However, this method requires that power system loads to be statistically independent, and assumes the need of metre measurement data. Differently, a singular value decomposition \cite{80966} technique  was employed in \cite{6996007subspace} to formulate a stealthy FDI attack using estimated subspace structure of measurements. 

In \cite{7001709Data-drivenPCA1}, in order to construct a blind stealthy FDI attack, a statistical model based on principal component analysis \cite{abdi2010principal} is used to transform the observed measurements into a linear combination of a vector of non-correlated principal components, which are the product of the Jacobian matrix of the power grid with a projected matrix. Their PCA-based blind cyberattack construction strategy has opened up a potential research direction and has been followed by range of academic researchers. However, the data-driven methods mentioned above are valid if the measurement matrix involves only AWGN. Adnan and Abdun \cite{7741557PCA3} \cite{ANWAR201758PCA4} have proven that in the case of gross errors, those blind attack strategies failed to pass the conventional BDD of the SE. In \cite{7741557PCA3}, the blind stealthy FDI attack is formulated based on matrix recovery problem by extracting the original low-rank measurement matrix and the gross error. Additionally, following similar principle to \cite{7741557PCA3}, in \cite{10.1007/978-3-319-31863-9_13} they formulate a data-driven undetectable malicious attack utilizing a low-rank and sparse matrix factorization methodology on the original measurement matrix with missing values. 
Finally, other recently suggested data-driven approaches include \cite{9026314PCA2}, \cite{7934033}, \cite{7885051} and \cite{8248780}.
\section{Classification Based on Attack Targets} \label{attTaerget}
Various cyber-physical elements are essential for monitoring and controlling the grid operation. However, they also make the Smart Grid vulnerable to a variety of data breaches that may bring a greater exposure to attacks on data integrity, data confidentiality, data availability, and so forth. FDI attacks target various cyber-physical components of the Smart Grid ranging across all domains, namely generation, transmission, distribution, consumption, market, and operations. In this sub-section, vulnerabilities of some of the principal cyber-physical elements are discussed. 
\subsubsection{EMS}
The EMS within the control center is the most affected target in smart power system. State estimator serves as an interface between the cyber space and the physical space, rendering it the most vulnerable element within the EMS of the Smart Grid environment. This is quite important, particularly because the processes within the EMS are temporally sequential. For example, the output of the SCADA or PMU systems are critically demanded by the state estimator, and the other subsequent EMS modules highly require the output of the state estimator. As a result, the state estimator is the most important target for cyber attackers. Coordinated and sophisticated cybertattacks, such as the FDI can compromise measurement data (targeting either the input to the SE or the outcome as discussed in \ref{reqFDI}). This can cause unbounded estimation errors and can deceive the system operator stealthily. Further, this can be seen from the various consequences of the FDI attacks against the SE as presented in \ref{attImpct}. 

Since the first paper \cite{liu2011false} of FDI attack, the majority of FDI attack methodologies described in \ref{attackModel} target the SE. The vulnerability issues in the SE problem can be investigated with respect to the various cyber-and physical elements, including Physical properties of the power system, communication systems, IEDs, and AMIs. Related attack targets also include transmission lines \cite{7271100topologyAtt5} \cite{8338154}, topology  \cite{7872498} \cite{6547837topologyAtt1} \cite{7185427topologyAtt2}, and system observability \cite{7286402}.
\subsubsection{Automation Generation Control}
In the power grid, data between AGC and generator units or NCSs is transmitted via communication systems such as SCADA and PMU, making them vulnerable to cyberattacks. Reference \cite {7286615AGC2} has experimentally evaluated that the AGC algorithm can be manipulated by adversaries on frequency measurements, generation of load balance, and control commands between AGC and generator units. Further, in \cite{6740883AGC3}, the authors studied data integrity attacks directed at the AGC. They defined various data integrity attack templates such as a scaling attack, ramp attack, pulse attack, random attack, and explored at how these attacks could modify the measurements and generator operating points through the AGC by providing an incorrect perception of the system load. However, instead of pursuing prescribed data integrity models, intelligent and coordinated adversarial models targeting AGC are likely to be tactical, and their strategies can be more adaptive during attacks. As a solution for this shortcoming, Tan et al \cite{tan2017modelingAGC4} are the first to research on the attack of false data on AGC's sensor measurements, demonstrating that FDI attacks on the power flow measurement vector can deceive grid frequency to reach certain safety-critical thresholds in the shortest possible time, without triggering at any integrity checks on the sensor data. A parallel line of research focusing on the FDI attack targeting the vulnerabilities of AGC and associated communication infrastructure can be found in \cite{khalaf2018joint} \cite{7573304}. 
\subsubsection{Contingency Analysis}
The feasibility of FDIs on CA through the SE is studied in \cite{8281015}. Attackers could stealthily introduce contingency of transmission line to a normal contingency list by misleading the CA process by injection of false data into the SE. The exploited contingency would then be embedded as security constraints in the security-constrained economic dispatch (SCED), which may result in various impacts (see Section \ref{lmp}). Similarly, FDI attack against the CA considering security constrained optimal power flow (SCOPF) and transmission line capacities is studied in \cite{rahman2019novel}. Accordingly, their findings have shown that potential FDI threat vectors could prevent CA such that the system can experience overloading conditions on one or more transmission lines when particular contingencies arise.  
\subsubsection{Distribution Energy Management}
Distribution energy management (DEM) \cite{6298960} has become so instrumental for handling real-time networks and dynamic decisions that could not otherwise be taken by conventional EMSs. More importantly, DEMs are highly applicable in distributed-based SEs and DERs/microgrids with the aim of maximizing the efficiency and quality of service in terms of minimizing outages, mitigating interruption time, and ensuring reasonable frequency and voltage levels \cite{SUN2005187}. Despite their popularity in the power grid, they face the unprecedented challenge from the incumbent cyberattack. The vulnerability of DEM to FDI was studied in \cite{6197400Distributed2}. It was found that the manipulated data introduced by the attacks would cause imbalanced demand and response, increase costs for electricity transmission and distribution, and affect the reliability of energy supplies in the power grid. The vulnerability of DEM to the false data attacks has been further explored with regard to dynamic microgrids, as demonstrated in reference \cite{7249479}.  
\subsubsection{Market Management System} \label{mms}
Market management system (MMS) \cite{ratnakaran2007market} is the national electricity market of the grid that dictates energy prices. MMS is designed to facilitate standardised transactions between service providers and utility consumers in the energy industry. The MMS provides market information based on variables such as price, dispatch and other constraints obtained from EMS/DEM modules such as SCOPF. Even so, MMS has become a primary goal for adversaries to manipulate intelligence on the utility market or otherwise to make illicit financial gains. Among the pioneer research works in FDI attacks against the MMS include \cite{5622048} \cite{6074981} \cite{6657769}. The financial risks induced as the result of such vulnerabilities are covered in Section \ref{lmp}.  
\subsubsection{Communication Systems}
Numerous communication technologies \cite{6298960} in the Smart Grid are vulnerable to the FDI attacks. Power system measurements are vulnerable to the FDI attacks, for instance via the SCADA \cite{7741557PCA3} system. This may further affect other cyber-physical elements such as the SE or AMI. In other words, if adversaries get access to the SCADA system they can damage AMI and the intruders can carry out falsifying customer billing information. Communication protocols (such as the IEC 61850) are also vulnerable to the FDI attack \cite{6298960}. Among the communication systems that can potentially bring vulnerability to the Smart Grid environment include NCS \cite{8270717}, WAMS \cite{wang2019two}, IEEE C37.118 \cite{6298960}, and wide area network communication infrastructure \cite{1213}.  
\subsubsection{Intelligent Electronic Devices}
IEDs link field devices to a communication infrastructure that enables SCADA and SAS to gather critical grid information. FDI attacks have been found to jeopardise such critical information by breaching IEDs \cite{7570183}. For example, FDI attacks can temper voltage readings over the IEDs, and they can modify IED settings that can also cause the relay to trip. This can also lead to an abrupt voltage drop below the critical level, resulting in load shedding and much worse, power outages.
\subsubsection{Renewable DERs}
DERs have been among the most vulnerable cyber-physical components to FDI attacks. In \cite{6197400Distributed2}, the vulnerabilities of DERs considering routing process have been investigated. Their discussion confirmed that the forged data injected by the attackers would induce imbalanced demand and response, cause higher costs for energy transmission, distribution, and the number of outage customers. 

Microgrids have become increasingly popular in the Smart Grid infrastructure owing to their versatility and integration with renewable energy. However, they have also become potentially susceptible to the exponentially escalating variety of cyber threats. In particular, their performance can worsen dramatically in the face of more intelligent FDI attacks. \cite{nikmehr2019game} \cite{7249479} \cite{7570183} \cite{8260848} are among the research efforts that examine vulnerabilities of microgrids to FDI attacks in the Smart Grid.
\begin{table}[h]
\caption{Impact of the FDI attacks in Smart Grid}
\centering
\begin{tabular}{m{2.5cm} m{6cm}}
  \Xhline{2\arrayrulewidth}
\textbf{Impact category} & \textbf{References} \\ \hline
Risk and reliability & \cite{wang2019two} \cite{6275516ACattack} \cite{7366616AC} \cite{khanna2017bi} \cite{basumallik2020impact} \cite{tan2015integrity}   \cite{8219710} \cite{8023151} \cite{7438904} \cite{7433442topologyAtt3} \cite{6148224} \cite{8468098} \cite{8338154} \cite{7470429} \\ \hline
Secure operation and stability & \cite{jiongcong2016impact} \cite{7249479} \cite{7286402} \cite{6148224} \cite{6848210}  \cite{7426376topologyAtt4} \cite{6740883AGC3} \\ \hline
Electricity market and pricing economics & \cite{6655273} \cite{8281015} \cite{6032057Centralised2} \cite{6074981} \cite{6831164}  \cite{8219710} \cite{8023151} \cite{5622048} \cite{7995087} \cite{6831166} \cite{7872498} \cite{6657769}  \cite{7440873} \cite{6522551} \cite{5607275} \cite{7109165} \cite{7401113} \cite{7903742} \cite{8905473} \cite{7004067} \\ \hline
Energy theft &  \cite{6655273} \cite{5622048} \cite{wang2014extended} \cite{RASHEDMOHASSEL2014473} \cite{ismail2020deep}  \\ \hline
Energy data privacy and confidentiality & \cite{RASHEDMOHASSEL2014473} \cite{bhattacharjee2017statistical} \cite{8254498}  \cite{7087397} \\ \Xhline{2\arrayrulewidth}
\end{tabular}
\label{impactAttack}
\end{table}
\section{Classification Based on Impact} \label{attImpct}
The study on the impact of FDI attacks across the electric power system has become one of the most interesting research direction. Therefore, it is important to quantitatively examine the possible severity of the physical or economic consequences of threats associated to the FDI. For example, if an adversary successfully launches an FDI attack that can control the results of the state estimator, the system operator can make non-optimal, uneconomic, or even dangerous power dispatch decisions on the results of the incorrect state estimate. Furthermore, discrepancies due to injection of malicious data in the SE can be amplified in the follow-up modules and lead to devastating consequences starting from tripping of a transmission line breakers or unsafe frequency fluctuations, to economic impacts, and blackouts in large geographic regions. In this sub-section, major impacts on the power grid, including secure operation and stability, risk and reliability, electricity market and pricing economics, energy theft, energy data privacy and confidentiality are presented.  Table \ref{impactAttack} summarises the related research papers. 
\subsubsection{Risk and Reliability}
Reliable supply of electricity is essential for any power system. Equally important, grid operators are expected to provide electricity to their customers at an acceptable risk level. In the mean time, the likelihood of cyber security events significantly impacts the reliability of power system. Cyber adversaries can have detail knowledge of the various cyber-physical components of the Smart Grid. This will help them to examine the cyber-to-physical mapping in the penetration of attack vectors that eventually impact the power system reliability \cite{4840054}. For instance, circuit breaker trips can be caused by the probabilities of successful cyber-capable attacks through the SCADA system \cite{6939397}, and through RTU \cite{8274710}. 

One of the major risks of FDI attacks is its ability to induce cascading failures. For example, attackers can intrude with injections of false data to deliberately cause overloaded branch trips \cite{8338154}, which can induce cascading failure and potentially do serious harm to power grids. In addition, adversaries can develop an optimal FDI attack to deliberately cause a re-dispatch \cite{7366616AC} of power generation that results in a physical overflow on the target transmission line,   and shutdown of a larger portion of the power grid \cite{6275516ACattack}. Moreover, LR attacks \cite{6148224} \cite{8468098} \cite{8338154} \cite{7470429} (see Section \ref{constMethod}) are some of the FDI attacks which have potential impacts on the reliability of power supply. For example, in \cite{7470429} the reliability of power system considering generator, line and load demand subject to the LR attack is evaluated. Finally, \cite{wang2019two} \cite{khanna2017bi} \cite{basumallik2020impact} \cite{tan2015integrity}  \cite{8219710} \cite{8023151} \cite{7438904} \cite{7433442topologyAtt3} are among some of the researches of FDI attacks that study the impacts of risk and reliability of Smart Grid infrastructure.
\subsubsection{Secure Operation and Stability}
When the power system is working under the range of acceptable limits it is known to be secure. Power system operators employ security assessment procedures, typically using static security assessment and dynamic security assessment to ensure the secure operation, system design, and stability of the power grid. Although a secure power system is engineered to tolerate contingency events, orchestrated hidden FDI attacks have catastrophic impact on the secure operation and the stability of the power system. The effect of FDI attacks against static security assessment was reported in \cite{jiongcong2016impact}. The authors considered two attack scenarios: fake secure signal attack and fake insecure signal attack. According to the finding, the former attack scenario misleads the control center to believe that the system works in a secure condition when it is not, and the latter attack scenario misleads the control center to take corrective actions, like generator rescheduling and load shedding when it is costly and unnecessary. Similarly, reference \cite{6740883AGC3} discussed the impact of data integrity attacks directed at the AGC on the stability of the power system and the operation of the electricity market. Similarly, the impact of FDI attack on real-time load measurement readings through AMI has been investigated \cite{6848210}. It was evaluated through a case study of load information modification for a load distribution and dispatch where the aim of the attacker is to inflict an instability to the power system by the sudden change in load. FDI attackers also impact the stability of microgrids, for instance, by falsely changing the measurements for energy supply and demand of consumers within the microgrids \cite{7249479}. \cite{jiongcong2016impact} \cite{7286402} \cite{6148224} \cite{7426376topologyAtt4} \cite{6740883AGC3} are some of the research works which have discussed consequential impacts on the secure operation and stability of the power grid. 
\subsubsection{Electricity Market and Pricing Economics} \label{lmp}
A successful FDI attack on the Smart Grid infrastructure would see serious economic impact. For example, a prolonged power outage as a result of the incumbent cyber threat can bring substantial economic losses within the grid and may further cause tremendous disturbances to other businesses that have dependency on the supply of electricity. The financial misconduct of cyberattackers through the FDI can be seen from two main perspectives: manipulation of electricity market and modification of loads via the economic dispatch in the EMS/MMS. These are explained below.

Stealthy FDI targeting EMS and MMS has an impact on power system operations, such as economic dispatch problem, a large-scale optimization problem in the Smart Grid, which aims to meet the system demand, at the lowest possible cost, subject to reliability constraints. The impact of FDI attack on the economic dispatch was demonstrated by the authors in \cite{8023151}, where they implemented an FDI attack model with full system knowledge against transmission line ratings to cause maximal congestion over critical lines, resulting in a breach of capacity limits. This illustrates the economic and safety risks raised by the use of the FDI exploited key parameters such as line ratings. 
 Power grid retailers charge for the electricity they supply to the market according to the locational marginal price (LMP) \cite{1198282} \cite{5622048} \cite{6074981} at their point of  connection to the system; and customers obtain the electricity they buy on the basis of the LMP at their point of connection. \cite{1198282} shows that day-ahead and real-time LMP algorithms utilise recurrent outputs of the SE. Consequently, the FDI-compromised SE outcomes have a significant impact on the electricity market, where falsified prices can be sent to customers. There are various literature that cover the impact of FDI attacks on the electricity market which are explained below. 

Xie et al \cite{5622048} were the first to show the impact of FDI attacks on the electricity market. Using a method considering Ex-Post market model for finding cases where price shift occurs, authors demonstrated the likely financial misconduct that can be triggered by the FDI cyberattacks while being undetected by the SE of the system operator. This line of research has been expanded to \cite{6074981}, which formalizes the economic loss due to the FDI attacks on real-time LMPs. They analysed the financial impact of FDI threats on energy market operations using day-ahead and ex-post real-time LMP models. They also suggested the likelihood that the malicious attack could give financial profit to the adversary by incorporating with virtual bidding. In \cite{6032057Centralised2}, the authors showed that an attacker can inject a malicious vector to change real-time and day-ahead market revenue of generation, and can potentially make a profit. The authors considered residue energy heuristic to determine especially harmful effects in weak attack regimes also showing various attack detection probabilities. Reference \cite{7440873} looks at the problem of FDI cyberattacks against the real-time pricing model that incorporates various DERs and traditional power resources. They considered the impacts of two attacks on the real-time pricing scheme: Ex-ante (FDI launched by the attacker before a decision-making process) and Ex-post (FDI launched by the attacker after a decision-making process). They analyzed a welfare gain and welfare loss with regard to the attack's impact on the real-time pricing system. In contrast to the previous studies, reference \cite{6522551} evaluated financial risk in electricity market operations, where the threat model was defined through inter-temporal constraints of an economic dispatch \cite{5607275}. 

Additionally, \cite{6831166} studied vulnerabilities of the electricity market through the LR attack, where the attacker can stimulate a false price of real-time electricity by constructing biased pattern of transmission congestion. Similarly, the impact of transmission line rating on electricity markets is studied in \cite{7109165}, where the real-time LMPs are exploited by falsified injections of transmission line rating vector. Further assumptions are made that the adversary has complete information of the system (including system load, generation cost information). Another research work on the consequences of FDI attacks on the real-time market operations is by the authors in \cite{6657769}, who modelled the real-time LMP using a geometric characterization to demonstrate the relationship between bad data and price. A similar research on the vulnerability of electricity market to the FDI attacks, \cite{7401113} considered a more practical adversarial model that could produce unpredictable pricing signal on the assumption of the attacker's incomplete knowledge of the power system. Other research works on the impact of FDI attacks in electricity market and pricing economics in Smart Grids include \cite{7903742}, \cite{8905473}, and \cite{7995087}. 

Unlike the above researches which are based on day-ahead, ex-post and ex-ante electricity market models, references \cite{7004067} and \cite{6831164} use multi-step electricity price (MEP) \cite{7004067} model, which has been implemented by many countries to encourage energy efficiency, load balancing, and fairness in energy consumption. The authors of \cite{7004067} proposed a two-dimensional MEP model to analyze and determine a desirable quantity and price of electricity in several steps, in which each step is scaled by both the time when the electricity is utilised and the quantity of electricity. As compared to the other electricity market models, MEP has been found to be robust against FDI attacks \cite{6831164}.  
\subsubsection{Energy Theft}
Energy theft is a growing concern that has incurred massive financial damages to electricity supply providers worldwide. There are different motives behind energy theft cyberattack using the FDI. For example, by manipulating a number of sensors and sending false measurement to the Regional Transmission Organizations, a malicious attackers aims to generate a profit from the market \cite{5622048}. As such, a stealthy injection of bad data can bring a profit to the adversary by exploiting a virtual bidding system. Another reason could be a malicious customer may exploit the electricity consumption computed by a smart meter to pay less than the actual value of the energy consumed.

Energy theft by an FDI attack has also been reported in \cite{wang2014extended}, which shows an attacker that minimises the measurement of active power on a standard bus power system by moving a power load from the $5^{th}$ bus to the $4^{th}$ bus. As a result, if the attack is successfully launched on the stated bus and the attack continues for one day, consumers connected to the $5^{th}$ bus may see their charges falsely reduced by \$272,871. Another case of energy theft via the FDI malicious hackers has occurred in AMIs, where attackers alter data of smart meter as it is transmitted over the network between the meter and the control center \cite{RASHEDMOHASSEL2014473}. Finally, in a very recent article \cite{ismail2020deep} it has been revealed that energy theft by malicious customers breach into the smart meters monitoring their renewable generation system and exploit their readings to demand higher energy supply to the national grid and thereby wrongly overcharge the utility provider.
\subsubsection{Energy Data Privacy and Confidentiality}
As well as it is a common understanding in data security, cyberattacks against data confidentiality put emphasis on a breach against data privacy of customers. Most of the studies on FDI attack impact are focused on the impacts mentioned above (such as energy theft and electricity market); however, privacy against customers in the grid emerges at various data monitoring interfaces, and hence needs special attention. In particular, smart meters act as unified interfaces between the cyber and physical environments of the Smart Grid, rendering them face risks from the combined cyber-physical attacks. Data flow between smart metres and utility centers include electricity usage activity and system monitoring commands. There are several ways coordinated attackers can have access to the smart meters. For example, they can bypass the cryptographic functions of smart meters, can have access to customer data, can manipulate it, and result in falsification of the data or even disordering integrity of the utility center. Additionally, the attackers can send mass packets to exhaust the bandwidth of communication of the smart meter and to further cause communication to disconnect \cite{7087397}. The attacker would then gain access to the data via a physical memory, and will execute unauthorised writing or reading operations in the physical memory. Again, when the communication is back to normal the newly injected or modified customer data is transmitted to the network. 

The forgery of the power consumption across the smart metre can be accomplished during either the collection of data (i.e. input to the metre) or during the transmission/reception in the AMI network \cite{bhattacharjee2017statistical}. Further, the data manipulation can happen when the data is at rest (i.e. storage of data within the meter) \cite{RASHEDMOHASSEL2014473}. Finally, the manipulated smart metres can expose customer's data, report a falsified power consumption data in the AMI, and could have substantial consequences on the operation of the Smart Grid.
\section{Literature Review Method}\label{revMethod}
A systematic search, selection, analysis, and critical evaluation of the literature is described in this section.
\begin{figure}[!ht]
	\centerline{\includegraphics[width=93mm]{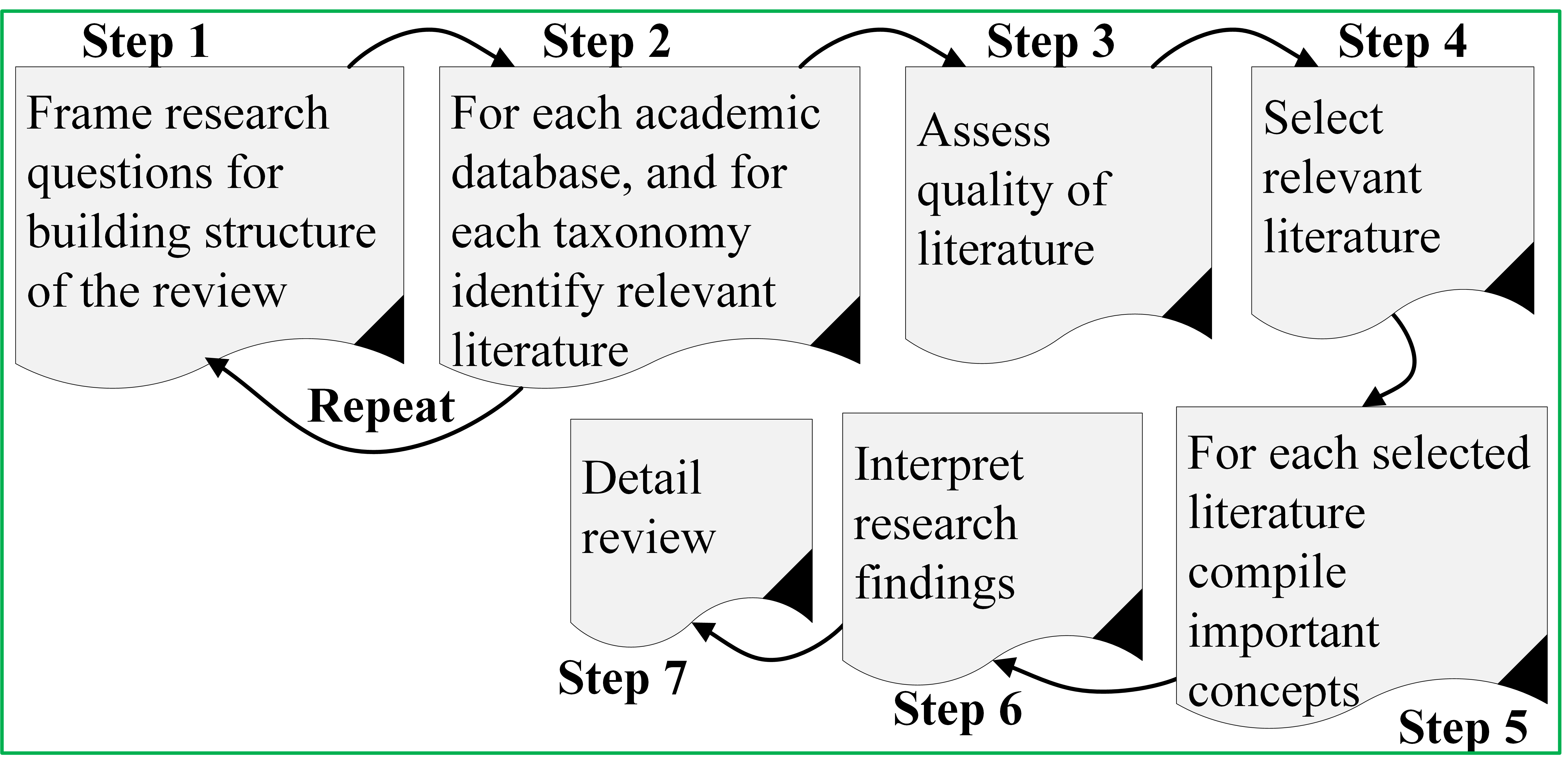}}
	\caption{Literature review methodology}
	\label{liteMethod}
\end{figure}
\subsection{Literature Search Methodology}
It seems that the literature search process plays an important role in crafting a comprehensive analysis of a topic. The literature survey of this paper is based on the search methodology adopted by Webster and Watson \cite{webster2002analyzing}. The systematic identification of high-quality publications (namely review articles, journals, conferences, and Books), technical reports, and dissertations are reflections of the correct selection of databases, keywords, the time covered, the papers considered in the literature search, and performing backward and forward searches \cite{alex213419}. 

Fig. \ref{liteMethod} is a description of the methodology used for literature search on this paper. The following academic research databases are considered: IEEE Xplore (IEEE/IET) digital library, Elsevier ScienceDirect, ACM digital library,  SpringerLink, and Others. To find relevant papers, Fig. \ref{liteMethod} is applied for each of the academic research databases. Using the first step, keywords using Google Scholar and Microsoft Academic were identified with respect to the adversarial model, attack targets, and impacts. "Smart Grid", "power system", "false data injection", and "cyber security" are common keywords used in each of the three classes. Accordingly, the following keywords were used for each class of the FDI attack (also using \textit{intitle}, \textit{AND}, \textit{OR} and other Google search engine advanced operators wherever necessary). 1)  \textbf{Adversarial model}: "Smart Grid", "power system", "cyber security", "false data injection", "adversary", "construction", "attack model". 2) \textbf{Attack targets}: "Smart Grid", "power system", "cyber security", "target", "vulnerable". 3) \textbf{Attack impacts}: "Smart Grid", "power system", "cyber security", "impact", "consequence", "effect".
\begin{table}[ht]
 \caption{Summary of relevant publications}
 \centering
 \begin{tabular}{m{1cm} m{0.75cm} m{1cm} m{1cm} m{0.5cm} m{1.5cm}} \Xhline{2\arrayrulewidth}
\textbf{Database source} & \textbf{Survey articles} & \textbf{Original res. articles} & \textbf{Conf. papers} & \textbf{Book} & \textbf{Tot. no. of relevant papers} \\  \hline
IEEE Xplore & 5 & 59& 19 & -&83 \\  \hline
Elsevier SD & 1 & 6  & - & - &7 \\ \hline
ACM & 1 &2 & 1 & -&4 \\ \hline
Springer &1  & 1  & - & 1 & 3 \\ \hline
Others & - & 3  & 1 & - & 4 \\ \hline
\textbf{Total} &\textbf{8} &\textbf{71} & \textbf{21} &\textbf{1}  & \textbf{101} \\ \Xhline{2\arrayrulewidth}
\end{tabular}
\label{pubLit}
\end{table}
\subsection{Literature Selection and Analysis}
Primarly, we reflect entirely on FDI threats with respect to the Smart Grid cybersecurity, as there are also FDI articles related to other areas such as WSN, healthcare, software-defined networks, and so on. Another consideration is, while all the scholarly research sources considered are prestigious and are assumed to publish quality works, further evaluations were made using scientific journal ranking platforms to assess quality of the journals and the CORE\footnote{CORE: Computing Research and Education Association of Australasia (https://www.core.edu.au/)} was used for the conferences. Based on the search method as described above, a systematic literature selection and analysis are used which are described here. First, aggressive search was conducted using the above keywords and Step 2 of Fig. \ref{liteMethod} that resulted in abundant number of papers. Then, after a systematic refinement across each taxonomy of the FDI, relevant literature was selected (Step 3 and Step 4 of Fig. \ref{liteMethod}. In addition to the keywords, titles and abstracts were considered for correctly categorising the selected papers. It also allowed us to subsequently re-categorize some literature as there were some publications that included more than one of the three classes. Next, important concepts were assembled for each of the chosen articles, accompanied by an overview of research results, and a thorough analysis (the last three steps of Fig. \ref{liteMethod}). After an in-depth analysis of the literature, approximately 101 papers are found which, to varying degrees, dealt with the topic of an FDI attack in Smart Grid cybersecurity under the three classes. Note that the study of FDI attack in Smart Grid started in the late 2009. Therefore, the search for the most relevant literature of our survey starts from 2009 up to December 31, 2020 although related literature such as the BDD goes back in time before 2009. Table \ref{pubLit} is a summary of the number and source of the relevant publications considered in our survey paper.     \begin{table}[ht]
 \caption{Evaluation criteria for the false data attacks in Smart Grid}
 \centering 
 \begin{tabular}{m{2cm} m{6cm}}
 \Xhline{2\arrayrulewidth}
 \textbf{Criterion} & \textbf{Description} \\ 
 \hline 
Attack model &  Review the cyberattacks from the point of considered adversarial construction model \\ \hline 
Approach & Review various technical approaches followed by researchers for the design of the attack models    \\ \hline
References &  Review which articles study which FDI attack model \\ \hline 
Network architecture &  Relevant articles are reviewed from network-centric point of view \\ \hline 
Power flow model &  Adversaries use different approaches with different power flow models, so the incumbent cyberattacks are reviewed and compared accordingly  \\ \hline
Attack target & Relevant articles are reviewed from the point of view  of vulnerabilities of cyber-physical system elements  \\ \hline
Attack impact & Articles are compared on the basis of risk of the cyberattack\\ \hline
Validation metric & Show the main claim of the research exemplifying the performance  \\ \hline
Experimental platform &  Show the theoretical proofs or hardware testbeds utilized to justify the method \\   \Xhline{2\arrayrulewidth}
    \end{tabular}
    \label{criteriaTable}
\end{table}
\begin{table*}
 \caption{Comparison of FDI attacks in Smart Grid cybersecurity}
 \label{cmpTable}
 \centering 
\begin{tabular}{|m{0.15cm}|m{1.15cm}|m{1.15cm}|m{0.1cm}m{0.1cm}m{0.1cm}m{0.1cm} m{0.8cm}m{0.1cm}m{0.1cm}|m{0.15cm}m{0.15cm}m{0.15cm}m{0.15cm}m{0.15cm}|m{3.5cm}|m{0.1cm}m{1.5cm}m{0.1cm}|}
 \hline
  & & &  \multicolumn{7}{c|}{\textbf{Attack target}} &  \multicolumn{5}{c|}{\textbf{Attack impact}}& & \multicolumn{3}{c|}{\textbf{Exp. platform}} \\      
 {\rotatebox{90}{\textbf{Attack model}}} & \textbf{Approach} &  \textbf{Reference} & {\rotatebox{90}{EMS}} & {\rotatebox{90}{AGC}} & {\rotatebox{90}{DEM}} & {\rotatebox{90}{MMS}} & {\rotatebox{90}{Network comm.}} & {\rotatebox{90}{Intelligent device}} & {\rotatebox{90}{Renewable DER}} 
 
& {\rotatebox{90}{Secure operation}} & {\rotatebox{90}{Risk and reliability}} & {\rotatebox{90}{Electricity market}} & {\rotatebox{90}{Energy theft}} & {\rotatebox{90}{Energy privacy}} 
 & \textbf{Performance metric} & {\rotatebox{90}{Simulation}}  &
  %{\rotatebox{90}{Economic and others}} & 
   {\rotatebox{90}{IEEE Bus system}} & {\rotatebox{90}{Test bed}} \\   \hline
{\multirow{50}{*} {\rotatebox{90}{\textbf{Complete topology information}}}} &  Heuristics &  \cite{liu2011false}$^{D,c}$ & $\checkmark$  &  &  & & SCADA &  &  &  & & & & & Prob. of attack vector vs \% of compromised meters & $\checkmark$ & 9, 14, 30, 118, 300 &    \\  \cline{3-19}  & &  \cite{sandberg2010security}$^{D,c}$ &$\checkmark$ &  &  & & SCADA, PMU & $\checkmark$& & & & & & & Security index bound vs measurement number &$\checkmark$ & 14 & \\\cline{3-19}
 & &  \cite{5464816Centralised4}$^{D,c}$ &$\checkmark$ &  &  & & $-$&  &  &  & & & & & ADR vs attack sparsity  &$\checkmark$ & 14 & \\  \cline{3-19}
&  & \cite{5751206Centralised3}$^{D,c}$ &$\checkmark$ &  &  &  & PMU&  &  &  & & & & &Prob. of sparsest attack vector vs \% of compromised meters & & 30, 57, 118, 300&  \\  \cline{2-19} 
&  Graph-theoretic
 &  \cite{6032057Centralised2}$^{D,c}$ & $\checkmark$&  &  & &PMU &  &  &  & & $\checkmark$ & & &DAR vs attack MSE, market revenue & $\checkmark$&14 & \\  \cline{3-19}
& &  \cite{6102319Distributed3}$^{D,d}$ &$\checkmark$ &  &  & &$-$ &  &  &  & & & & & ADR vs FPR& $\checkmark$& $-$&  \\  \cline{3-19}
& &  \cite{6275516ACattack}$^{A/D,c}$ &$\checkmark$ & &  & &SCADA &  &  &  & $\checkmark$&  & & & Injected meas vs line number. \# of compromised RTUs vs line/bus.&$\checkmark$ &57 &
\\ \cline{3-19}
& &  \cite{6939486}$^{A/D,c}$ &$\checkmark$ & &  & $\checkmark$&SCADA &  &  &  & &  & & & \% of attack vs SR &$\checkmark$ &57 & 
\\  \cline{3-19}
& &  \cite{8425789ACincompleteInfo2}$^{A,d}$  &$\checkmark$ &  &  & & SCADA&  &  &  & & & & &Prob. of successful attack vs system information completeness & $\checkmark$& 9, 14, 30, 118, 300& \\  \cline{3-19}
& &  \cite{8625609}$^{D,d}$  &$\checkmark$ &  &$\checkmark$  & & SCADA&  &  &  & & & & &System states vs bus number; \# of compromised measurements vs \# of attacked states & $\checkmark$& 13, 37& 
\\  \cline{3-19}
& &  \cite{7249479}$^{d}$  & &  & $\checkmark$ & & Smart meter& & $\checkmark$ & $\checkmark$ & & & & &Energy loss (KWH) vs \# of supply units & $\checkmark$& 30&

\\  \cline{3-19}
& &  \cite{wang2014extended}$^{A,d}$& & & $\checkmark$ & & Smart meter& &  & $\checkmark$ & $\checkmark$& & & &Attack success probability vs injection level & $\checkmark$& 14, 39, 118, 300&
\\  \cline{2-19} 
&  LP
 &  \cite{tan2017modelingAGC4}$^{D,c}$  & & $\checkmark$ &  & & SCADA &  &  &  & & & & & Frequency deviation vs AGC cycle index, Compromised power flow vs AGC cycle index&$\checkmark$ &16 &$\checkmark$ \\ \cline{3-19}
&   & \cite{5947717}$^{D,c}$  &$\checkmark$ &  & & &SCADA, PMU &  &  &  & &$\checkmark$ & & &Real-time revenue vs detection probability, attack target locations & $\checkmark$& 14& \\  \cline{2-19}

&  LASSO & \cite{6486001Distributed4}$^{D,c/d}$ &$\checkmark$ &  & $\checkmark$ & & PMU &  &  &  & & & & &Prob. of attack vector vs SR vs & $\checkmark$& 9, 57&  \\  \cline{3-19}
 &   & \cite{ozay2013sparse}$^{D,c/d}$  &$\checkmark$ &  &$\checkmark$  & &PMU &  &  &  & & & & &Prob. of attack vector vs SR vs & $\checkmark$& 9, 30, 57, 118& \\  \cline{2-19}
&  SDP
&  \cite{8403288AC}$^{A,c}$  &$\checkmark$ &  &  & & SCADA, PMU &  &  &  & & & & & Spurious values vs original values of measurements; attack sparsity vs regularisation & $\checkmark$&30 & \\ \cline{2-19}
&  Bi-level MILP & \cite{7366616AC}$^{A,c}$ & $\checkmark$&  &  & &SCADA &  &  &  & $\checkmark$& & & &Prob. of attack vector vs load shift constraints & $\checkmark$& 24& \\ \cline{3-19}  &&  \cite{khanna2017bi}$^{D,c}$ & $\checkmark$& $\checkmark$ & & &SCADA & & & &$\checkmark$ & & & &Generator and line contingency vs attack vector & $\checkmark$& 14, 30& \\  \cline{3-19}
& & \cite{choeum2019oltc}$^{A,d}$ &  &  & $\checkmark$ & &AMI, smart meter &  &  & $\checkmark$ & & & & &  Compromised system states vs \# of buses& $\checkmark$& 33 &  \\  \cline{3-19}
& & \cite{8219710}$^{D,c}$ & $\checkmark$ &  &  & &$-$ &  &  & & $\checkmark$ & $\checkmark$ & & &  Financial benefit (\$/hour) vs total load (MW)& $\checkmark$& 14, 30 &  \\ \cline{2-19}
&  MINLP & \cite{8281015}$^{D,c}$ & $\checkmark$& $\checkmark$ &  & &SCADA &  &  &  & & $\checkmark$ & & &LMP vs bus number; LMP deviation vs attack cases & $\checkmark$& 14& \\ \cline{2-19} 
& Differential Evolution  & \cite{jiongcong2016impact}$^{D,C}$ & $\checkmark$&  & &  & SCADA &  & $\checkmark$ & $\checkmark$ & & $\checkmark$& & & System states vs measurement number  & $\checkmark$& 39 & \\ \cline{2-19} 
& Multi-objective Opt.  & \cite{6831164} & $\checkmark$&  & &  & Smart meter &  &  &  & &$\checkmark$ & & & Prices vs FDI attack cases  & $\checkmark$& 39 & \\ \cline{2-19} 
& Game-theoretic  & \cite{8270717} & $\checkmark$&  & &  & PMU, NCS &  &  &  & & & & & Adversarial cost  vs defence budget  & $\checkmark$& $-$ & \\  \cline{3-19}
& & \cite{nikmehr2019game}$^{d}$ &  &  & $\checkmark$ & &PMU &  &  $\checkmark$& $\checkmark$ & & & & &  Adversarial cost  vs defence budget; Power mismatch vs time& $\checkmark$& $-$ &\\  \cline{3-19}
& &  \cite{7570183}$^{d}$  & &  & $\checkmark$ & & $-$& $\checkmark$ & $\checkmark$ &  & & & & &Load shading cost (MW) vs \# of attack round & $\checkmark$& 9, 14& \\  \cline{3-19}
& & \cite{8023151}$^{D,c}$ & $\checkmark$ &  &  & &$-$ &  &  &  & $\checkmark$& $\checkmark$& & & Attacker's optimal gain (line cap. violation), manipulated line rating (MW) vs time (hour) & $\checkmark$& 118 & $\checkmark$ \\ \cline{2-19} 
& Clustering  & \cite{6655273} & & & $\checkmark$ &  & SCADA &  &  &  & & $\checkmark$& $\checkmark$& & Compromised measurement vs injection attack & $\checkmark$& 14 &
\\ \hline \end{tabular}
\end{table*} 
\begin{table*}
 \centering 
%\scalebox{0.85}{
\begin{tabular}{|m{0.15cm}|m{1.15cm}|m{1.15cm}|m{0.1cm}m{0.1cm}m{0.1cm}m{0.1cm} m{0.8cm}m{0.1cm}m{0.1cm}|m{0.15cm}m{0.15cm}m{0.15cm}m{0.15cm}m{0.15cm}|m{3.5cm}|m{0.1cm}m{1.5cm}m{0.1cm}|}
 \hline
  & & &  \multicolumn{7}{c|}{\textbf{Attack target}} &  \multicolumn{5}{c|}{\textbf{Attack impact}}& & \multicolumn{3}{c|}{\textbf{Exp. platform}} \\      
 {\rotatebox{90}{\textbf{Attack model}}} & \textbf{Approach} &  \textbf{Reference} & {\rotatebox{90}{EMS}} & {\rotatebox{90}{AGC}} & {\rotatebox{90}{DEM}} & {\rotatebox{90}{MMS}} & {\rotatebox{90}{Network comm.}} & {\rotatebox{90}{Intelligent device}} & {\rotatebox{90}{Renewable DER}} 
 
& {\rotatebox{90}{Secure operation}} & {\rotatebox{90}{Risk and reliability}} & {\rotatebox{90}{Electricity market}} & {\rotatebox{90}{Energy theft}} & {\rotatebox{90}{Energy privacy}} 
 & \textbf{Performance metric} & {\rotatebox{90}{Simulation}}  &
  %{\rotatebox{90}{Economic and others}} & 
   {\rotatebox{90}{IEEE Bus system}} & {\rotatebox{90}{Test bed}} \\   \hline
{\multirow{16}{*} {\rotatebox{90}{\textbf{Partial topology information}}}} &  KICA &  \cite{8581440ACincompleteInfo3}$^{D,c}$ & $\checkmark$ &  &  & &SCADA &  &  &  & & & & & Time of attack construction vs degree of incomplete info; Prob. of attack vector vs \% of incomplete info & $\checkmark$ & 14, 30, 118 &  \\  \cline{2-19}
&  Bi-level MILP 
 & \cite{7741928}$^{D/A,c}$ &  & $\checkmark$ &  & & $-$ &  &  &  & & & & & System re-dispatch vs SR & $\checkmark$ & 24 & \\  \cline{3-19}
& & \cite{8323244}$^{D,c}$ & $\checkmark$ &  &  & &$-$ &  &  &  & & & & & Physical PF vs injected PF & $\checkmark$& 24, 118 &  \\\cline{2-19}
&  Heuristics & \cite{7438904}$^{D,d}$ &  & $\checkmark$ &  & & $-$&  &  &  & $\checkmark$ & & & &  Line outages vs load measurement attack & $\checkmark$& 6, 96 &    \\  \cline{3-19}
& & \cite{7401110ACincompleteInfo1}$^{A,c}$ &  & $\checkmark$ &  & &PMU &  &  &  & &$\checkmark$ & & & Attack cost vs attacking regions & $\checkmark$& 14, 118 & \\  \cline{2-19}
&  Graph-theoretic &  \cite{8260948}$^{A,d}$ & $\checkmark$ &  &  & & SCADA, PMU &  &  &  & & & & &Measurement residual vs \% of estimation error of attacked state variables  & $\checkmark$& 30, 118 & \\  \cline{2-19}
&  Semi-Markov Process &  \cite{basumallik2020impact}$^{D,c}$ & $\checkmark$ &  &  & & PMU, IEC 61850 &  &  & $\checkmark$ & $\checkmark$ &$\checkmark$ & & & Attack probability, risk index, impact (generator/line loss)  & $\checkmark$& 200, 500 & \\  \cline{2-19} 
&  $-$ &  \cite{9285056}$^{A,c}$ & $\checkmark$ & $\checkmark$ &  & & $-$ &  &  & $\checkmark$ & &$\checkmark$ & & & Generation schedule, system congestion vs malicious load vector  & $\checkmark$& 30 & 
\\  \cline{2-19} 
&  $-$ &  \cite{6672638AC1}$^{A,c}$ & $\checkmark$ & &  & & PMU &  &  &  & & & & & Change in residue vs measurement number & $\checkmark$& 30 &\\  \cline{2-19} 
& RTP  &  \cite{6074981}$^{D,c}$ & $\checkmark$ &  &  &$\checkmark$ & SCADA &  &  &  & &$\checkmark$ & & & Real-time pricing (RTP) vs bus location  & $\checkmark$& 14 & \\  \cline{3-19}  &  &  \cite{tan2015integrity}$^{D,c}$ & $\checkmark$ &  &  & & SCADA &  &  & $\checkmark$ &$\checkmark$ &$\checkmark$ & & & RTP, stability vs attack vector  & $\checkmark$& 14, 30 &$\checkmark$ \\ \cline{2-19} 
& SDP  & \cite{7995087}$^{D,c}$ & &  & $\checkmark$ &  & SCADA &  &  &  & & $\checkmark$& & &  Profit confidence vs attack undetectability, information uncertainty  & $\checkmark$& 14, 118 &
\\ \cline{2-19} & Game-theoretic  & \cite{7446354}$^{D,c}$ &$\checkmark$ &  &  &  & PMU &  &  &  & & $\checkmark$& & &  Defender's loss vs \# of attackers; LMP (in MWH) vs bus number  & $\checkmark$& 30 &
\\  \cline{3-19}
& & \cite{wang2019two}$^{A,c}$ & $\checkmark$ &  &  & &PMU &  &  & & $\checkmark$& & & &  Load shading value (in MW)  vs attacked lines, attack-defence strategy& $\checkmark$&14, 57, 118  & \\ \hline

{\multirow{5}{*} {\rotatebox{90}{\textbf{LR attack}}}} 
&  Heuristics& \cite{6831166}$^{D,c}$ & $\checkmark$ &  &  & & SCADA & &  &  & &$\checkmark$ & & & Real-time LMP vs bus number, dispatch interval & $\checkmark$& 6 &    \\  \cline{2-19}

&  Bi-level MILP & \cite{yuan2011modeling}$^{D,d}$ & $\checkmark$ &  &  & & SCADA & $\checkmark$ &  &  & &$\checkmark$ & & & Generation dispatch, economic loss vs attack quantity & $\checkmark$& 14 &    \\  \cline{3-19}
& & \cite{7286402}$^{D,d}$ & $\checkmark$ &  &  & &$-$ &  &  & $\checkmark$ & & & & &  Load sensitivity vs attack resource& $\checkmark$& 14 &  \\  \cline{3-19}
& & \cite{6148224}$^{D,d}$ & $\checkmark$ &  &  & &SCADA &  &  & $\checkmark$ &$\checkmark$ & $\checkmark$& & &  Generation dispatch, economic loss, operation cost vs attack quantity& $\checkmark$& 14 & \\  \cline{3-19}
& & \cite{6805238}$^{D,d}$ & $\checkmark$ &  &  & &SCADA &  &  &  & & & & &  Power flows vs load attack magnitude& $\checkmark$& 14 & \\  \cline{3-19}
& & \cite{8468098}$^{D,c}$ & $\checkmark$ &  &  & &SCADA, PMU &  &  &  & $\checkmark$& & & &  Load level (in MW) vs load attack magnitude& $\checkmark$& 118 &

\\  \cline{3-19} & & \cite{8338154}$^{D,c}$ & $\checkmark$ &  &  & &SCADA &  &  &  & $\checkmark$& & & &  Load reduction (\%) vs lines tripped & $\checkmark$& 118 &
\\ \cline{2-19}

&  Tri-level MILP &  \cite{7271100topologyAtt5}$^{D,d}$ &  & $\checkmark$ &  & &$-$ &  &  &  & & $\checkmark$ & & & Optimal dispatch plan (in MW) vs injected attack & $\checkmark$& 14 & \\\cline{3-19}
& & \cite{7031948DCincomleteInfo3}$^{D,d}$ & $\checkmark$ &  &  & &SCADA &  &  &  & & & & & Percentage of attacking regions & $\checkmark$& 24, 30. 39, 57, 118 & \\  \cline{2-19}
&  Semi-Markov Process & \cite{7470429}  &$\checkmark$ &  & & &SCADA &  &  &  & $\checkmark$& & & &Load curtailment, $P_d$ vs \# of attacked substations &$\checkmark$ &79 & \\  \cline{2-19}
&  Game-theoretic & \cite{xiang2017game} &$\checkmark$ &  & & &SCADA & $\checkmark$ &  &  & & & & &Load curtailment vs system state number, defence budget &$\checkmark$ &24 &
\\  \cline{2-19}
&  Graph-theoretic & \cite{lee2019vulnerability}$^{D,c}$ &$\checkmark$ &  & & &SCADA, Router & &  &  & & & & &Load ratio  vs \# of removed nodes &$\checkmark$ &39 & \\  \cline{3-19}
& & \cite{6848210}$^{A/D,d}$ & &  & & $\checkmark$&AMI, Smart meter & & $\checkmark$ & $\checkmark$ & & & & &Meter current and power flow vs time &$\checkmark$ &17 &$\checkmark$ \\  \cline{3-19}
&  & \cite{8260848}$^{A,d}$ && & &$\checkmark$ &$-$ & & $\checkmark$ &  & & & & &Compromised system states  vs time &$\checkmark$ &4 & 
\\ \hline \end{tabular}
\end{table*} 

%\scalebox{0.9}{
\begin{table*}
 \centering 
\begin{tabular}{|m{0.15cm}|m{1.2cm}|m{1.2cm}|m{0.1cm}m{0.1cm}m{0.1cm}m{0.1cm} m{0.7cm}m{0.1cm}m{0.1cm}|m{0.15cm}m{0.15cm}m{0.15cm}m{0.15cm}m{0.15cm}|m{3cm}|m{0.1cm}m{1.5cm}m{0.1cm}|}
 \hline
  & & &  \multicolumn{7}{c|}{\textbf{Attack target}} &  \multicolumn{5}{c|}{\textbf{Attack impact}}& & \multicolumn{3}{c|}{\textbf{Exp. platform}} \\      
 {\rotatebox{90}{\textbf{Attack model}}} & \textbf{Approach} &  \textbf{Reference} & {\rotatebox{90}{EMS}} & {\rotatebox{90}{AGC}} & {\rotatebox{90}{DEM}} & {\rotatebox{90}{MMS}} & {\rotatebox{90}{Network comm.}} & {\rotatebox{90}{Intelligent device}} & {\rotatebox{90}{Renewable DER}} 
 
& {\rotatebox{90}{Secure operation}} & {\rotatebox{90}{Risk and reliability}} & {\rotatebox{90}{Electricity market}} & {\rotatebox{90}{Energy theft}} & {\rotatebox{90}{Energy privacy}} 
 & \textbf{Performance metric} & {\rotatebox{90}{Simulation}}  &
  %{\rotatebox{90}{Economic and others}} & 
   {\rotatebox{90}{IEEE Bus system}} & {\rotatebox{90}{Test bed}} \\   \hline
{\multirow{3}{*} {\rotatebox{90}{\textbf{GT attack}}}} &  Graph theoretic & \cite{6547837topologyAtt1}$^{D,c}$  &$\checkmark$ &  & & &SCADA &  &  &  & & $\checkmark$& & &DAR vs target branch, congested lines vs real-time LMP &$\checkmark$ &14, 118 & \\ \cline{3-19}
& & \cite{7450168}$^{A,c}$ &$\checkmark$ &  & & &SCADA &  &  &  & & & & & $P_d$ vs target line &$\checkmark$ & 24 &
\\ \cline{3-19}
& & \cite{chung2018local}$^{A,c}$ &$\checkmark$ &  & & &SCADA &  &  &  & & $\checkmark$ & & & Transmission loss (in MVA) vs line number &$\checkmark$ & 14, 118 & \\  \cline{2-19}
&  Heuristics & \cite{7438904}$^{D,d}$ & $\checkmark$ &  &  & &$-$ &  &  &  &$\checkmark$ & & & & Attack budget vs impact severity &$\checkmark$ & 96 &  \\  \cline{3-19}
& & \cite{7426376topologyAtt4}$^{D,d}$ &$\checkmark$  &  &  & &SCADA &  &  & $\checkmark$ & & & & &Line outages vs load measurement attack  & $\checkmark$& 14, 24, 30, 39, 57, 118 & \\ \cline{2-19} 
& Metaheuristics & \cite{liang2017framework}$^{A,c}$ &$\checkmark$ & & & &$-$ &  &  &  & & & & &Economic loss vs relative perturbation factor (\%); relative perturbation factor (\%) vs target line  & $\checkmark$& 39 &
\\  \cline{2-19}
&  Bi-level MILP & \cite{7185427topologyAtt2}$^{D,d}$ & $\checkmark$ &  & & &SCADA &  &  & $\checkmark$ & & & & & Line outages vs load measurement attack &$\checkmark$ & 14, 118 &  \\  \cline{3-19}
& & \cite{7433442topologyAtt3}$^{D,d}$ &$\checkmark$  &  &  & & PMU &  &  &  & $\checkmark$ & & & & Line outages vs load measurement attack & $\checkmark$& 39, 118 & \\ \hline 
{\multirow{6}{*} {\rotatebox{90}{\textbf{Data-driven}}}} &  ICA & \cite{7307141}$^{D,c}$ & $\checkmark$& & & &SCADA &  &  &  & &$\checkmark$ & & & MSE of ICA vs SNR, \# of observations; LMP vs bus& $\checkmark$& 14, 30& \\ \cline{2-19}
&  PARAFAC & \cite{7885051}$^{D,c}$ & $\checkmark$ & & & & SCADA& & & & & & & & MSE vs \# of intercepted meters; $P_{md}$ vs $\tau$ & $\checkmark$&14, 30 &
\\  \cline{2-19}
&  PCA & \cite{6996007subspace}$^{D,c}$ & $\checkmark$ & & & & SCADA& & & & & & & & Normalised SE error (\%) vs attack magnitude; ADR vs attack magnitude & $\checkmark$&14, 118 &  \\ \cline{3-19}
& & \cite{7001709Data-drivenPCA1}$^{D,c}$ & $\checkmark$ & & & & SCADA& & & & & & & & $P_{md}$ vs $\tau$ & $\checkmark$&14 & \\ \cline{3-19}
& & \cite{7741557PCA3}$^{D,c}$ & $\checkmark$ & & & & SCADA& & & & & & & & $P_{md}$ vs $\tau$ & $\checkmark$&14 &  \\ \cline{3-19}
& & \cite{10.1007/978-3-319-31863-9_13}$^{D,c}$ & $\checkmark$ & & & & SCADA& & & & & & & & Measurement residue vs \# of observations & $\checkmark$&14 &
\\ \cline{3-19}
& & \cite{ANWAR201758PCA4}$^{D,c}$ & $\checkmark$ & & & & SCADA, PMU& & & & & & & & Compromised measurements vs \# of observations; Compromised states vs \# of state variables; $P_{md}$ vs $\tau$& $\checkmark$&14, 30, 57 &
\\ \cline{2-19}
&  Geometric  & \cite{7934033}$^{D/A,c}$ & $\checkmark$ & & & & SCADA& & & $\checkmark$ & & & & & $P_{md}$ vs $\tau$, SCED operation cost vs malicious load & $\checkmark$&14, 30 & \\ \cline{2-19}
&  $-$  & \cite{8307441Distributed1}$^{A,c}$ & $\checkmark$ & & & & PMU & & & & & & & & Measurement residue vs \# of observations & $\checkmark$&56 & \\ \cline{2-19}
&  POMDP  & \cite{8248780}$^{A,d}$ & $\checkmark$ & & & & IEC 61850 & & &$\checkmark$ & & & & & Voltage sag vs attacked bus, Attacked bus vs time & $\checkmark$&39, 118 &
\\ \cline{2-19}
&  Deep RL  & \cite{9094321}$^{A,c}$ & $\checkmark$ & & & & PMU & & & & & & & & Load measurement (in MWA) vs  bus number; Attack resources vs training episodes & $\checkmark$&30 & \\ \cline{2-19}
&  Eliminate-Infer-Determine   & \cite{tian2018data}$^{D,c}$ & $\checkmark$ & & & & Smart meter & & & & & & & & Attack vector vs bus number; $P_{md}$ vs $\tau$ & $\checkmark$&14, 30, 118, 300 &
\\ \cline{2-19}
&  Random matrix theory  & \cite{lakshminarayana2020data}$^{D,c}$ & $\checkmark$ & & & & $-$ & & & & & & & & Attack sparsity vs $P_d$& $\checkmark$&118 &
\\  \hline \end{tabular}
\begin{tablenotes}
\item $[Ref^{D/A}]$: DC/AC model, $[Ref^{c/d}]$: centralised/decentralised architecture, $[Ref^{cd}]$: centralised and decentralised architectures, $[Ref^{RL}]$: real load data considered, ADR: attack detection rate, DD: Detection delay, FPDR: False positive DR, DA: Detection accuracy, FPR: False positive rate, TPR: True positive rate, FDI: injected magnitude of FDI attack, payoffs: Game metric of attacker-defender cost in payoffs, SR: FDI attack sparsity ratio, SNR: Signal-to-noise ratio, MAPE: Mean absolute percentage error, PE: Percentage error between true and estimated states, AR: attacking rate (Attackability, or successful attacking probabilities), MSE: Mean square error, $P_d$: Probability of detection, $P_{md}$: Probability of missed detection, PARAFAC: PARallel FACtor analysis, $\tau$: attack detection decision threshold, POMDP: Partial Observable Markov Decision Process, MINLP: Mixed-Integer Nonlinear Programming, NFP: Nonlinear Fractional Programming, SDP: Semidefinite Programming.  \end{tablenotes}
\end{table*}
\subsection{Evaluation Criteria}
In order to quantify the efficacy and associated challenges of the different cyberattack strategies, several key evaluation criteria are suggested in relation to the requirements of the power systems and the Smart Grid cybersecurity. The assessment criteria used to compare the selected false injection attack papers are summarized in Table \ref{criteriaTable}.

The evaluation criteria are used to compare and
 contrast among the various attack construction methods, attack targets and impacts as detailed in Section \ref{taxoAll}, \ref{attTaerget}, \ref{attImpct} respectively, and summarised in Table \ref{cmpTable}. One of the main evaluation criteria is attack model, a criterion that reflects the reviewed FDI threat construction model. Five commonly used attack construction methodologies have been considered for the attack model criterion, namely attack with complete information, attack with partial information, LR attack, GT attack, and attack using data-driven. The other evaluation criterion is approach or algorithm for the design of the attack models. The various approaches for the evaluation of the literature mentioned in this survey paper include heuristics, meta-heuristics, graph-theoretic, game-theoretic, bi-level \& tri-level Mixed-Integer Linear Programming (MILP), Statistical transformation approaches (PCA, ICA, PARAFAC), Markov and ML models. Furthermore, the AC and DC models are considered for the power flow model. The reviewed articles are also evaluated from network-centric point of view (considering centralised and decentralised architecture). Note that the power flow model and the network architecture are used as super-script of the 'Reference' (column 3 Table \ref{criteriaTable}). Most importantly, the FDI attack papers are investigated with regards to the attack target and attack impact
 evaluation criterion. Notice that the different components of the Smart Grid can be seen from
the discussion in Section \ref{backG}. Finally, two evaluation criteria, namely performance metrics and experimental platform have been inspected.  
\section{Comparison and Statistics Among Defence Strategies}\label{comparisonSec}
In our review paper, 101 publications are considered for the three classes of the false injection attacks. Here, the various strategies are compared and some statistical facts based on the evaluation criteria are presented. 
\subsection{Adversarial Model}
Since the original conception of the FDI attacks by Liu \cite{liu2011false}, most adversarial models have been assumed by the full knowledge of the underlying power system operations. Accordingly, this category of adversarial model comprises around 42\% of the total works surveyed. These adversarial models with full knowledge of network data and topological settings have been on the premise that attackers could hack any more of the stringent power system security controls. The adversarial model with limited knowledge of topological and network data is more reasonable than the adversarial model with complete knowledge that makes the other most popular approach used in the Smart Grid cybersecurity community. In this case, approximately 18\% of the total surveyed publications have come within this threat model. Adversarial models leveraging data-driven approaches are relatively the latest and are the second most popular research areas with respect to FDI attack construction strategies at present, standing at about a fifth of the total surveyed literature. Notably with the emergence of cyber-physical datasets, these strategies are more appealing in the handling of the complex Smart Grid infrastructure. The other well researched attack models are LR attack and GT attack, which are very harmful and have very serious consequences, as described in Section \ref{attackModel}. Both of these attack types account for a fifth of the total surveyed publications.
\subsection{Attack Target}
Although many of the IT and OT elements of the Smart Grid are vulnerable to the cyberattack, EMS and SCADA/PMU are found to be the most vulnerable control and monitoring systems. This is due to the fact cyber attackers aim to compromise the SCADA measurement data or try to manipulate the outcome of the EMS/DEM (Refer Section \ref{attTaerget}). Most of the FDI attacks consider EMS and SCADA/PMU as the main target elements, accounting for almost 95\% of the other critical elements. Consequently, other key OT elements such as the AGC, economic dispatch, and MMS will also be at a greater risk. Some attackers also try to compromise the sensor data via the IEDs/RTUs, communication systems such as the AMI, IEC 61850, DNP3, and Modbus, and. As compared to other sub-domains, the vulnerability issues of renewable DERs and microgrids have got little attention. 
\subsection{Attack Impact}
Details of the investigation into the impact of cybersecurity attacks on the cyber-physical systems are shown in Section \ref{attImpct}. Almost half of the surveyed articles examined the impact (directly or indirectly) of the FDI attack on the Smart Grid. In fact, one third of these studies are related to the impact of the FDI attack on the economic dispatch and electricity market. Moreover, around 30\% of the surveyed papers analysed the effect of FDI attacks on the secure operation and power system reliability. Finally, a relatively limited number of papers (just under 10\%) looked at the impact of  the incumbent cyberattack on energy theft and customer data privacy.
\subsection{Performance Metric}
The FDI attacks vary, among other things, in terms of the construction model, algorithmic design, attack target, and network architecture. For this reason, instead of providing a distinct performance metrics for all the adversarial models, we present comprehensive qualitative metrics. A plentiful of performance metrics are presented for each of the countermeasure subcategories (see $16^{th}$ column of Table \ref{cmpTable}). For example, across the complete topology information category, optimal/subset of meter/IED protection, attack cost are the main metrics considered. Further, packet loss, computational cost, communication cost, and end-to-end delay are the main evaluation metrics adopted among the prevention schemes. In most of the detection based on dynamic SE, statistical-based models, and data-driven defence categories, detection rates (in terms of probability of detection, TPR) are compared against false positive rates or false alarm rates.   
\subsection{Experimental Platform}
The vast majority of studies performed numerical results based on simulations of IEEE standard or modified electric grid test cases. Various sizes of test cases have been considered, IEEE 14 bus system being the most widely referred test case. Although the vast majority of literature use only a single test case to conform their numerical results, some considered multiple test cases. The majority of the studies are based on simulations using MATPOWER\footnote{https://matpower.org/} optimization toolbox. To further verify the efficacy of their proposal just a very few of the scholars incorporated a real-time testbed. 
\section{Main Gaps of Existing False Data Attack Researches} \label{litGap}
In what follows, we describe the key gaps of existing FDI attack researches.

\textbf{Some Emerging Smart Grid Areas Are Not Well Studied}:
The plethora of literature examined in this review paper tried to cover a multitude of Smart Grid infrastructures; however, there are some open issues with respect to the scope (network architecture, DERs, and communication systems). The majority of existing cyberattack researches have focused on the traditional centralised EMS. Hence, FDI threat models against distribution systems of SE is still an open research. For example, adversarial construction methodology in realistic multi-phase and unbalanced smart distribution systems and DERs \cite{8625609} can be more interesting. While decentralized energy generation and distribution systems (such as the DERs) have become very popular, yet they can be among the most vulnerable cyber-physical components to the orchestrated FDI attacks. But, only few research studies have been undertaken with respect to the attack construction and/or impact of the flase data attacks against the DERs. This can be seen from the $15^{th}$ column of Table \ref{cmpTable}. Further, only few papers have discussed FDI attacks in the SAS, AMI, and WAMS-based  communication systems. Our survey also reveals that the impact of FDI attack on energy theft and user data privacy is another research area with just little attention at the moment. \\

\textbf{Need for Further of FDI Attack for AC-Based Systems}:
Most existing FDI attack experiments are performed in a confined setting on the assumption that the functions from the power system states to the measurements are linear (DC-based power flow models). Although this approach can be a very good assumption, many industry standard SE models are of non-linear AC power flow. Therefore, although the AC power flow model is far more complex than the DC counterpart, cybersecurity practitioners and other stakeholders need to come up with the stealthy FDI attack for industry-wide AC-based SEs. It would be more interesting, if the cyberattack can be explored in large-scale realistic EMS/DEM applications considering industrial-based AC state estimators that involve dynamic contingency analysis.\\

\textbf{Need for Corroboration of Experimental Results Via Testbed Platform}:
Although the literature surveyed in this paper have proven their cybersecurity solutions via numerical simulations benchmarked against standardised test cases, it is vital to validate the experimental results via cyber-physical testbeds, which is missing in the literature except to a few of them (\cite{8555556} \cite{8248780} \cite{7822933}).
This downside can be seen from the perspectives of data- and system-oriented approaches. Most of the FDI attack schemes surveyed did not consider commercial-level datasets, which otherwise, can practically validate the vulnerability of the state estimators to the stealthy FDI attacks.   

Testbeds \cite{6473865} are essential tools for testing the performance evaluation of algorithms and protocols in the Smart Grid. The highly complex and multidisciplinary essence of the Smart Grid requires the implementation of cyber-physical testbeds with different characteristics for comprehensive experimental validation. There is a considerable need to analyse new Smart Grid security concepts, architectures, and vulnerabilities via cyber-physical system test platforms. More recently, there has been a growing attention to the study of cyber-physical Smart Grid testbeds \cite{6473865}. Most notably,  hardware-in-the-loop test platforms have become much more popular for the development, analysis, and testing of cyber-physical components of the electrical power system. For example, some Smart Grid stakeholders, such as ABB\footnote{https://new.abb.com/news/detail/62430/abbs-acs6000-power-electronics-grid-simulator-pegs-tests-medium-voltage-equipment}, Siemens Power Technologies\footnote{https://assets.new.siemens.com/siemens/assets/api/uuid:1fb8264a-9ee6-4d71-a703-bb68beb7ca94/version:1587982708/rtds-datasheet-en-1909.pdf}, and OPAL RT\footnote{https://www.opal-rt.com/hardware-in-the-loop/} foster hardware-in-the-loop testing using real-time digital simulators across various Smart Grid realms, including microgrids, SAS- and WAMS-based protection environments. Therefore, we suggest that assessing the effects of FDI attacks on the Smart Grid using the hardware-in-the-loop testbed platform is critical in crafting the stringent cybersecurity requirements. 
\section{Emerging Advanced Applications: Future Research Directions} \label{futurePros}
Securing the electricity grid is one of the highest priorities of many countries around the world. Academic studies and industries are expected to tackle a range of issues for future research on cybersecurity attacks in the Smart Grid infrastructure. Particularly, the reliance of reliable and secure power system operation on the communication infrastructure, along with potential cyber threats are increasingly growing. In the following, potential emerging advanced applications are discussed as means of future research prospects.\\

\textbf{Cybersecurity for Emerging Smart Grid Communication Systems}: 
Despite the fact that the communication infrastructure is the most critical target to the FDI attacks, the study of threat modeling and impacts have to be studied well, especially, across the SAS-compliant IEC 61850 and the WAMS-compliant IEEE C37.118. The FDI attack can well be studied especially with the incorporation of cyber-physical testbed platforms \cite{6473865}. Moreover, although AMI is one of the most vulnerable communication systems to the FDI attack, little has been done the risks associated with the incumbent cyberattack. Especially, given the increasing adoption of WSN and IoT in the Smart Grid, it will be interesting to address cybersecurity issues of IoT-based AMI with regard to the FDI attacks.

Software-defined networking is one of the emerging networking applications. The coupling of software-defined networking with the Smart Grid applications can bring efficient network monitoring. However, the security issue of this technology is worth investigating especially with respect to the FDI attacks.
Further, FDI attacks on heterogeneous cognitive radio, WSN and IoT are potential cybersecurity researches which are worth investigating. The application of data-driven models across the more intelligent communication arena of the Smart Grid can be explored to tackling against the orchestrated cyberattacks.\\

\textbf{Security Framework Based on Lightweight ML}: Countless memory and computational-restricted wireless sensor nodes are connected to IoT applications in Smart Grid. Several reports have shown that such limitations raise obstacles to the usage of conventional security measures over IoT systems. For example, from a defence against the FDI attack perspective, security frameworks using lightweight ML \cite{sliwa2020limits} can be proposed for resource-constrained IoT devices. On top of that, lightweight ML can be proposed for prevention schemes such as encryption, message authentication, and dynamic key management against the false data attacks in an end-to-end Smart Grid communication system.\\   

\textbf{FDI Attack in Edge Computing}: In a distributed computing environment, edge computing \cite{diro2018distributed} improves the communication overhead and system bandwidth by bringing the processing and data storage near to the origin of data source. Further, the emergence of Industry 4.0 \cite{foukalas2020cognitive} across a number of industries, including the Smart Grid, brings ubiquitous networked elements, and intelligent edge computing. While edge computing provides considerable advantages, it can also lead attackers with an easy point of entry to some of the cyber-physical edge devices that can then be used to obtain access to the core components of the Smart Grid. For instance, bringing more IoT devices to the edge network can introduce various cybersecurity threats like the FDI. Hence, the FDI attack is worth investigating across edge computing-based Smart Grid.\\  

\textbf{Distributed Electricity Trading}: 
The prevalence of DERs promotes the concept of distributed electric energy. Distributed electricity trading is one of emerging applications for a device-to-device energy sharing. As such the vulnerability and comprehensive risks of FDI attacks against LMP market pricing can be investigated in regards to DEM applications.\\

\textbf{Blockchain Technology}: 
As an innovative distributed computing ecosystem, Blockchain offers a secure solution for facilitating the immensely complex interactions among various cyber-physical Smart Grid entities. False injection attack across the Blockchain ecosystem is a very new research area, which requires a further investigation (for instance, privacy preservation and anomaly detection).
\section{Conclusion}\label{cncln}
Smart Grid poses a rising threat from an emerging cyber-physical attack called FDI.  By injecting falsified attack vectors stealthily, adversaries can violate the availability, integrity, and confidentiality of critical Smart Grid data, and may render the power system unobservable. In addition, coordinated FDI attacks can pose serious consequences for the Smart Grid, such as causing sequential transmission line outages, maximizing operation cost of the power system, culminating in large-scale failure of the power system operation, and regional/national catastrophic impacts.

This survey paper analysed the FDI attacks in Smart Grid in three main classes, namely the attack model, the attack target, and the attack impact. In order to quantify the efficacy and associated challenges of the various cyberattack models in the literature surveyed, a number of key evaluation criteria were used in relation to the requirements of the power systems and the Smart Grid cybersecurity. Finally, future research directions for FDI attacks are also proposed as a way of advancing the Smart Grid cybersecurity framework.

\bibliographystyle{srt}
\bibliography{Haftu_latex.bbl}
\end{document}